\documentclass[traditabstract]{aa}

\usepackage{epsfig,caption}
\usepackage{graphics}
\usepackage[latin1]{inputenc}
\usepackage{natbib}

\begin{document}
\title{
Variability selected high-redshift quasars on SDSS Stripe 82
}
\author{N. Palanque-Delabrouille\inst{1} \and Ch. Yeche\inst{1} \and A. D. Myers\inst{2,6} \and P. Petitjean\inst{3} \and Nicholas P. Ross\inst{4} \and E. Sheldon\inst{5} \and E. Aubourg \inst{1,7} \and T. Delubac\inst{1} \and J.-M. Le Goff\inst{1} \and I. Pâris\inst{3} \and  J. Rich\inst{1}  \and K. S. Dawson\inst{9}  \and D. P. Schneider\inst{10 }\and B. A. Weaver \inst{8}  
}
\institute{CEA, Centre de Saclay, Irfu/SPP,  F-91191 Gif-sur-Yvette, France 
\and  Department of Astronomy, University of Illinois at Urbana-Champaign,
 Urbana IL 61801, USA
\and  Universit\'e Paris 6, Institut d'Astrophysique de Paris, CNRS UMR7095, 98bis Boulevard Arago, F-75014 Paris, France 
\and Lawrence Berkeley National Lab, 1 Cyclotron Road, Berkeley, CA 94720, USA 
\and Brookhaven National Laboratory, Bldg 510, Upton, NY  11973, USA 
\and Max-Planck-Institut f\"ur Astronomie, K\"onigstuhl 17, D-69117
 Heidelberg, Germany 
 \and  APC, 10 rue Alice Domon et L\'eonie Duquet, F-75205 Paris Cedex 13, France 
 \and Center for Cosmology and Particle Physics, New York University, New York, NY 10003 USA 
 \and University of Utah, Dept. of Physics \& Astronomy, 115 S 1400 E, Salt Lake City, UT 84112, USA 
 \and Department of Astronomy and Astrophysics, The Pennsylvania State University, 525 Davey Laboratory, University Park, PA 16802, USA
}
\date{Received xx; accepted xx}
\authorrunning{N. Palanque-Delabrouille et al.}
\titlerunning{Variability selected high-redshift quasars on SDSS Stripe 82}
\abstract{
The SDSS-III BOSS Quasar survey will attempt to observe $z>2.15$ quasars at a density of at least 15 per square degree to yield  the first measurement of the Baryon Acoustic Oscillations in the Ly-$\alpha$ forest. To help reaching this goal, we have developed a method to identify quasars based on their variability in the $u g r i z$ optical bands. 
The method has been applied to the selection of quasar targets in the SDSS region known as Stripe 82 (the Southern equatorial stripe), where numerous photometric observations are available over a 10-year baseline. This area was 
observed  by BOSS during September and October 2010. 
Only 8\% of the objects selected via variability are not quasars, while 90\% of the previously identified high-redshift quasar population is recovered. 
The method allows for a significant increase in the $z>2.15$ quasar density over previous  strategies based on optical ($ugriz$) colors, achieving a density of 24.0~deg$^{-2}$ on average 
down to $g\sim 22$ over the 220~deg$^2$ area of Stripe 82. 
We applied this method to simulated data from the Palomar Transient Factory and 
from Pan-STARRS, and showed that even with data that have sparser time sampling than what is available in Stripe 82, 
including variability in future quasar selection strategies would lead to increased target selection efficiency in the $z>2.15$ redshift range. We also found that 
Broad Absorption Line quasars are preferentially present in a variability than in a color selection. 
}
\maketitle

\section{Introduction}
Baryonic Acoustic Oscillations (BAO) and their imprint on the matter power
spectrum were first observed in the distribution of galaxies~\citep{bib:cole,bib:BAO}. 
They can also be studied by using the H{\sc i} Lyman-$\alpha$ absorption signature of
the matter density field along quasar lines of sight~\citep{bib:white03, bib:McDonald07}. A measurement
sufficiently accurate to provide useful cosmological constraints requires the
observation of at least $10^5$ quasars, in the redshift range $2.2 < z < 3.5$, over
at least 8000 $\rm deg^2$~\cite{bib:eisenstein11}. This goal is one of the aims of the Baryon Oscillation Spectroscopic
Survey (BOSS) project~\citep{bib:schlegel09}, part of the Sloan Digital Sky Survey-III\footnote{http://www.sdss3.org} which is currently taking data. One of the challenges of this survey is to build a list of targets that contains a sufficient number of quasars in the 
required redshift range. 

Quasars are traditionally selected photometrically, based on their colors in various bands \citep{bib:schmidt83, bib:croom01, bib:richards04, bib:richards09, bib:croom09}. While these methods achieve good completeness at low redshift ($z<2$), they  present serious drawbacks for the selection of quasars at redshifts above 2.2. In particular, as was shown in~\cite{bib:Fan}, quasars with $2.5<z<3.0$ tend to occupy the same region of optical color space as the much more numerous stellar population, causing the selection efficiency (or purity) to drop below $\sim 50\%$ in that region. The same confusion occurs again for $3.3<z<3.8$. 
This was recently confirmed by \cite{bib:worseck} who have demonstrated that the SDSS standard 
quasar selection systematically misses quasars with redshifts in the range $3<  z< 3.5$. 

The separation of stars and quasars in the redshift range of interest can be improved 
by using the variability of quasars in the optical bands.
Light curves sampled every few days over several years were used by the MACHO collaboration~\citep{bib:MACHO}  
to identify 47 quasars behind the Magellanic Clouds. In a similar way, the OGLE project~\citep{bib:OGLE} has identified 5 quasars behind the Small Magellanic Cloud. Three seasons of observation on high galactic 
latitude fields were used by QUEST to search for variable sources. Nine previously unknown 
quasars~\citep{bib:QUEST} were discovered. 

More recently, significant progress in describing the evolution with time 
of quasar fluxes has been made possible by the multi-epoch data in the SDSS Stripe 82~\citep{bib:york00}. 
Using large samples of over 10,000 quasars, \cite{bib:deVries04} and \cite{bib:McLeod08} have 
characterized quasar light curves with structure functions. 
Concentrating on SDSS Stripe 82 data, 
\cite{bib:schmidt10} developed a technique for selecting quasars based on their variability. Recent 
works have shown that the optical variability of quasars could be related to a continuous time stochastic process driven by thermal fluctuations~\citep{bib:brandon09} and modelled as a damped random 
walk~\citep{bib:McLeod_a, bib:kozlowski}.  This resulted in a structure function that was used by~\cite{bib:McLeod_b} 
to separate quasars from other variable point sources. A variant, based on a statistical description of the variability in quasar light curves, was suggested by~\cite{bib:butler10} for the selection of quasars using time-series observations in a single passband. 

In this paper, we present a method  to select quasar candidates, inspired from the formalism developed by
\cite{bib:schmidt10}. The method was adopted by the BOSS collaboration to choose the objects that 
were targeted, during September and October 2010, in Stripe 82. 
This region covers 220~deg$^2$ defined by 
equatorial coordinates $ -43^\circ < \alpha_{\rm J2000} < 45^\circ$ and $-1.25^\circ <  \delta_{\rm J2000}   < 1.25^\circ$. It was previously imaged about once to three times a year from 2000 to 2005 (SDSS-I), then with an increased cadence of ~10-20  times a year from 2005 to 2008 (SDSS-II) as part of the SDSS-II  supernovae survey~\citep{bib:frieman08}.
With a sampling of 53 epochs on average, over a time span of 5 to 10 years~\citep{bib:abazajian09}, the SDSS Stripe 82 data are ideal for testing a variability selection method for quasars. For the first time, in September and October 2010,  
the observational strategy of BOSS rested entirely on variability for the final selection (after loose initial color cuts as explained below). In contrast, all target lists 
in BOSS had been obtained so far from the location of the objects in color-color diagrams, following various strategies --- such as the kernel density estimation  method~\citep{bib:richards04} or 
a neural network approach~\citep{bib:yeche10}.

Section~\ref{sec:method} presents the formalism used to describe the variability in quasar light curves and 
gives the performance of the chosen selection algorithm on quasar and star samples. Section~\ref{sec:targetting} explains how this tool was applied to select two sets of targets
in Stripe 82, and presents the results obtained. An extrapolation of this method to the 
full 10,000~deg$^2$ observed by SDSS, made possible by adding data from the Palomar Transient Factory~\citep{bib:PTF}, or 
from Pan-STARRS~\footnote{http://pan-starrs.ifa.hawaii.edu/public/home.html}, is presented in Section~\ref{sec:PTF_PS1}. We conclude in Section~\ref{sec:conclusions}.

\section{Variability selection algorithm}\label{sec:method}

The main purpose of this study was to develop an algorithm to select quasars in Stripe 82 based on their variability, while  rejecting as many stars as possible. Spectroscopically confirmed stars and quasars in Stripe 82 were used to compute two sets of discriminating variables. The first one, used to distinguish variable objects from non-variable stars,  consists in the $\chi^2$ of the  light curve with respect to the mean flux, in each of the five photometric  bands. The second one, which helps discriminating quasars from variable stars, consists in  parameters that describe the structure function.

\subsection{Quasar and star samples}\label{sec:samples}
We describe below the two samples, one of stars and one of quasars, which are used to test the variability algorithms, and to train the neural network of Sec.~\ref{sec:NN}.

\begin{figure}[h]
\begin{center}
\epsfig{figure=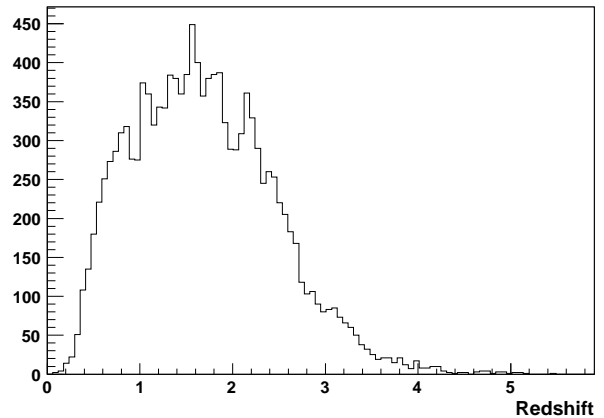,width = \columnwidth} 
\caption[]{Redshift distribution of the sample of quasars from all previous quasar surveys covering Stripe 82. } 
\label{fig:z_distrib_known}
\end{center}
\end{figure}

For the quasar training sample,  we used a list of 13328 
spectroscopically confirmed quasars obtained from the 2dF quasar catalog 
\citep{bib:croom04}, the 2dF-SDSS LRG and Quasar Survey (2SLAQ)~\citep{bib:croom09}, the SDSS-DR7 spectroscopic database~\citep{bib:abazajian09}, the SDSS-DR7 quasar catalog~\citep{bib:schneider10} and the first year of BOSS observations. These quasars have redshifts 
in the range $  0.05 \leq z  \leq 5.0 $ (cf. Fig~\ref{fig:z_distrib_known}) and $g$ magnitudes  in the  range
$18 \leq g  \leq 23$ (Galactic extinction-corrected). 

For the star sample, we used 2697 objects observed by BOSS, initially
tagged as potential quasars from color selection and spectroscopically confirmed as stars. Variability and color-selection are not fully independent: bright objects that are easily discarded by their colors are also easier to discard by their variability. Therefore, the use of these spectroscopically confirmed stars constitutes a conservative approach and corresponds exactly to the type of objects that we want to reject with the variability algorithm. 

\begin{figure}[h]
\begin{center}
\epsfig{figure=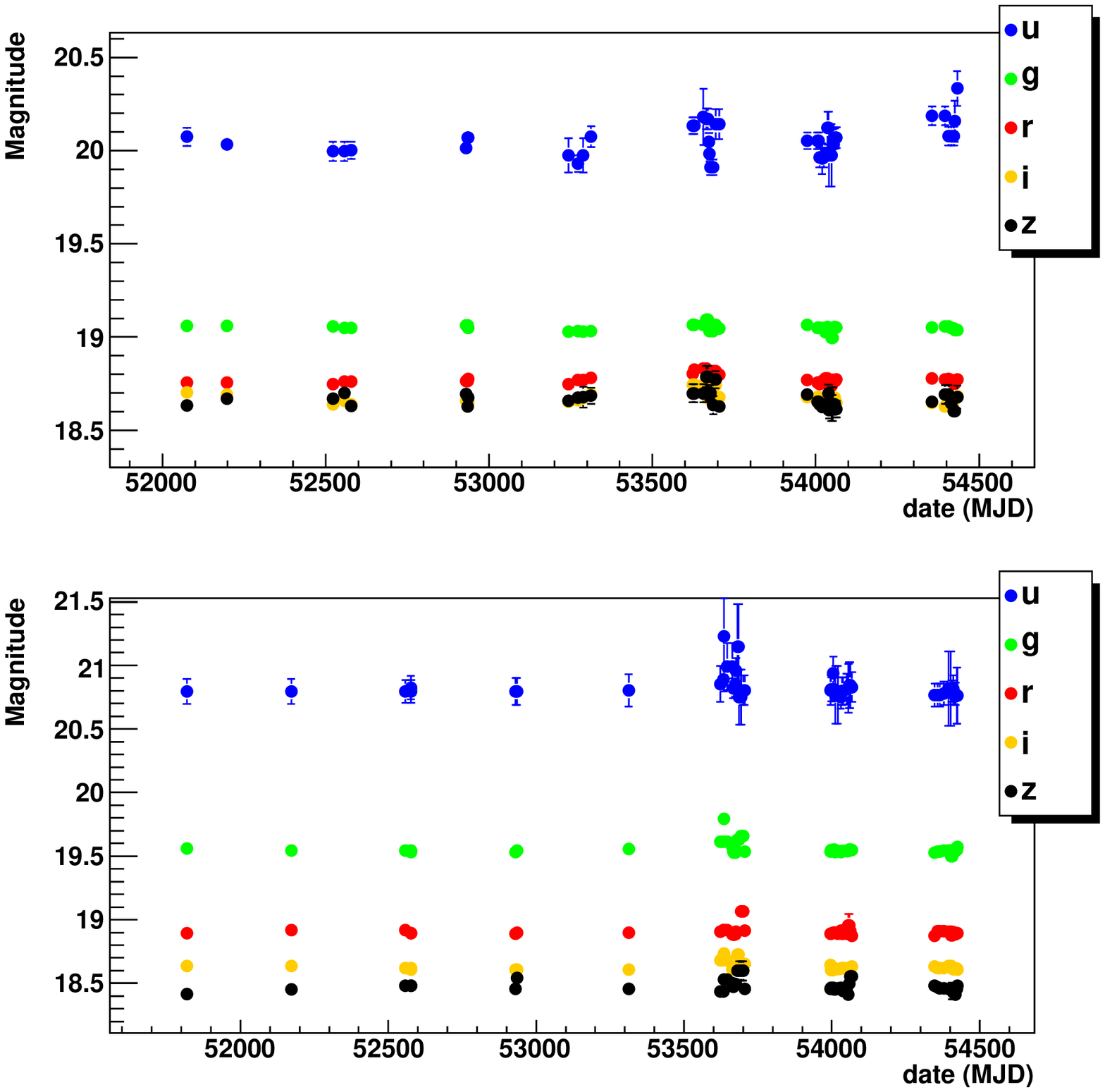,width = \columnwidth} 
\caption[]{Examples of  light curves (after median filtering and clipping as explained in Sec.~\ref{sec:LCcleaning}) in the five SDSS photometric bands for stars in Stripe 82.} 
\label{fig:LCstar}
\epsfig{figure=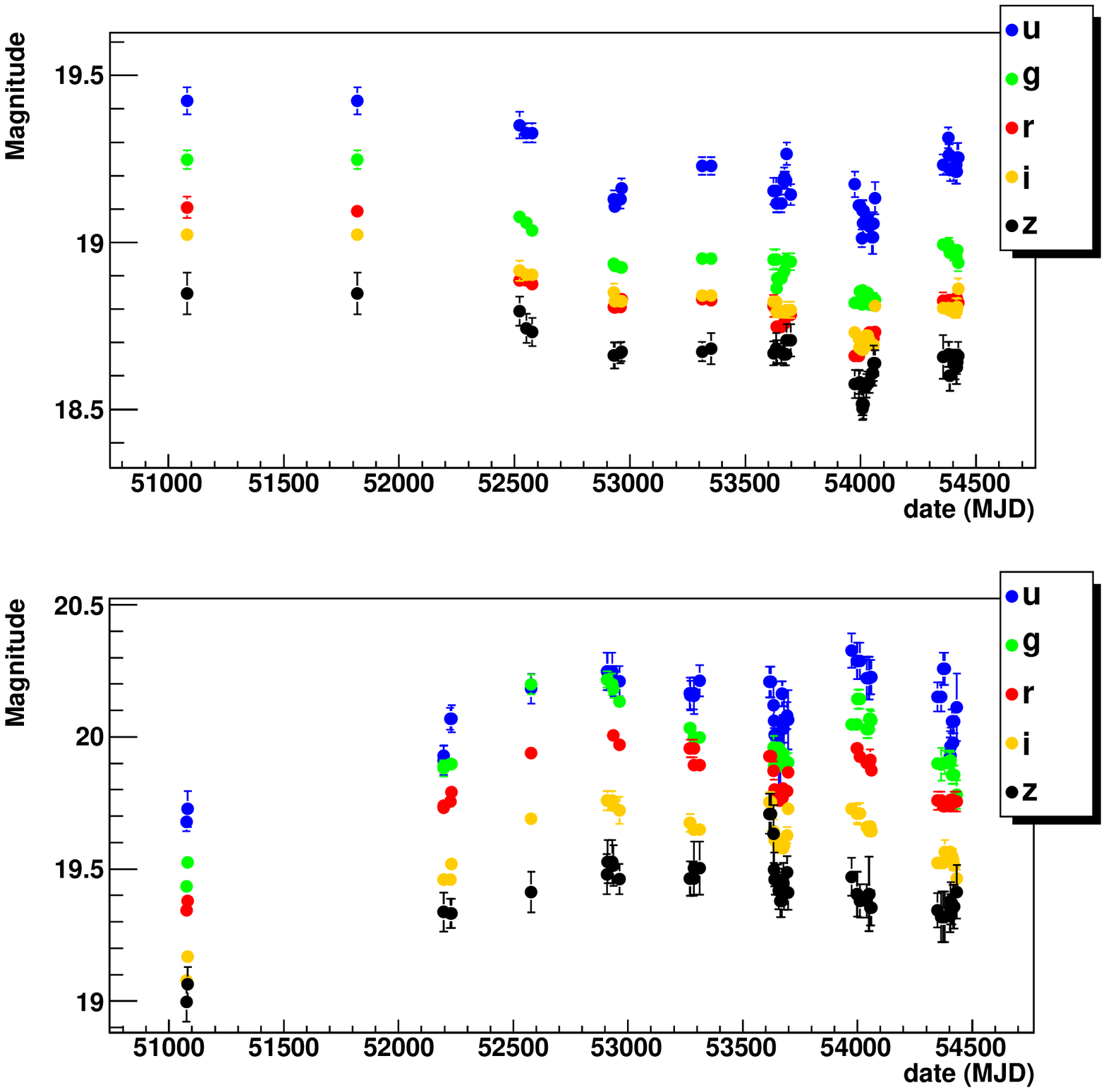,width = \columnwidth} 
\caption[]{Examples of  light curves  (after median filtering and clipping as explained in Sec.~\ref{sec:LCcleaning}) in the five SDSS  photometric bands for quasars in Stripe 82. } 
\label{fig:LCqso}
\end{center}
\end{figure}

Light curves were constructed for these two samples from the data collected by SDSS. The collaboration used the dedicated Sloan Foundation 2.5-m telescope~\citep{bib:gunn06}. A mosaic CCD camera~\citep{bib:gunn98} imaged the sky in five $ugriz$ bandpasses~\citep{bib:fukugita96}. The imaging data were processed through a series of pipelines~\citep{bib:stoughton02} which performed astrometric calibration, photometric reduction and photometric calibration.
Typical examples of stellar and quasar light curves are shown in Figs.~\ref{fig:LCstar} and \ref{fig:LCqso} respectively. The increased cadence after MJD 53500 are the SDSS-II supernovae search observations.

 The star and quasar samples have similar time samplings, representative of the typical time sampling on Stripe 82 (cf. Figs.~\ref{fig:LCstar} and \ref{fig:LCqso}). The number of epochs (i.e. number of photometric measurements in a given band)  varies from 1 to 140, with a mean of 53 and a r.m.s. of 20. The time lag between the first and the last epochs is  8  to 10 years long for 74\% of the targets, between 5 and 7 years long for 24\% and at most 4 years long for the remaining 2\%.
For this study, we concentrated on objects with at least 4 observation epochs, independently of the timespan. As a result, all targets that meet this requirement (13063 spectroscopically confirmed quasars and 2609 stars) have observations spanning at least two consecutive years.

\subsection{Pre-treatment of the light curves}\label{sec:LCcleaning}
Photometric outliers could alter significantly the values of the variability parameters, to the point of washing out any relevant information. The raw light curves were therefore cleaned of deviant points (irrespective of their origin, whether technical or photometric) in a two-step procedure. A 3-point median filter was first applied to the full quasar light curve in each of the five bands, followed by a clipping of all points that still deviated significantly from a fifth order polynomial fitted to the light curve. Note that to avoid removing too many photometric epochs, the clipping threshold, initially set at $5\sigma$, was iteratively increased until no more than 10\% of the points were rejected.  Despite the poorer frequency of the SDSS-I measurements (compared to SDSS-II), the median filtering was applied to the full light curve as the variations looked for are expected to occur on periods of several years. 

\subsection{Light curves $\chi^2$}\label{sec:LCchi2}
While most stars have
constant flux, quasars usually exhibit flux variations. As shown by~\cite{bib:sesar07}, at least 90\% of bright quasars are variable at the 0.03 mag level, and the variations in brightness are on the order of 10\% on time scales of months to years~\citep{bib:vandenberk04}.

\begin{figure}[h]
\begin{center}
\epsfig{figure=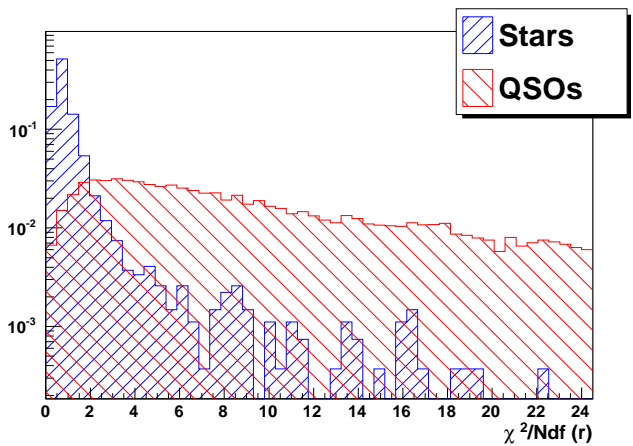,width = \columnwidth} 
\caption[]{Normalized distribution of the reduced $\chi^2$ in the $r$ band that results from fitting the light curves by a constant, for the stellar (blue) and the quasar (red) test samples. As confirmed by their larger reduced 
$\chi^2$, quasars clearly exhibit much larger deviations from a constant flux than stars. } 
\label{fig:chi2}
\end{center}
\end{figure}

Each of the $ugriz$ light curves were fit by a constant flux, and the resulting $\chi^2$ recorded. While most stars have a reduced $\chi^2$ near unity, as expected for non-variable objects, quasar light curves tend to be
poorly fit by a constant, resulting in a large reduced $\chi^2$, as illustrated in Fig.~\ref{fig:chi2} for the $r$ band. The $\chi^2$ thus helps to distinguish non-varying stars from varying point sources. 

\subsection{Variability structure function}\label{sec:structurefct}
\begin{figure}[h]
\begin{center}
\epsfig{figure=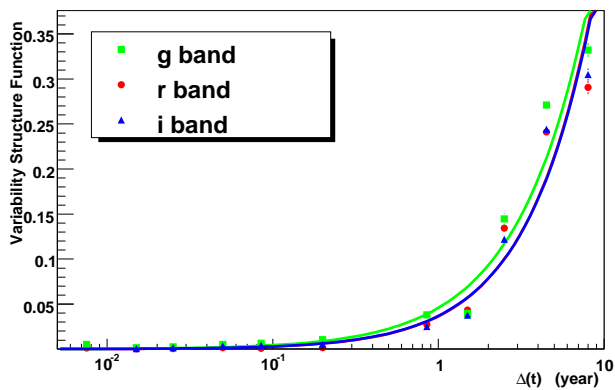,width = \columnwidth} 
\caption[]{Variability structure function $\mathcal V(\Delta t) $ of equation~\ref{eq:SF}  for a typical quasar. The curves show the best-fit power law $A\,(\Delta t)^\gamma$ for the three bands $g$, $r$, $i$. Note that the $r$ and $i$ best-fits are almost identical. } 
\label{fig:SF}
\end{center}
\end{figure}

The structure function characterizes light curve variability by quantifying the change in amplitude $\Delta m_{\rm ij}$ as a function of time lag $\Delta t_{\rm ij}$ between observations at epochs $i$ and $j$. Following the prescription of~\cite{bib:schmidt10}, the variability structure function of the source magnitude, is given by
\begin{equation}
\mathcal V(\Delta t_{\rm ij}) = |\Delta m_{\rm i,j}| - \sqrt{\sigma_{\rm i}^2 + \sigma_{\rm j}^2}\, , \label{eq:SF}
\end{equation}
where $\sigma$ is the magnitude measurement error.
The structure function can be modeled by a power law $A\,(\Delta t)^\gamma$ in all photometric bands, with $\gamma>0$, illustrating the fact that, for quasars,
the r.m.s. of the distribution of the magnitude difference between two observations tends to increase with time lag (cf. Fig.~\ref{fig:SF}). 

To derive the power law parameters $A$ and $\gamma$ for a given light curve, we define the likelihood 
\begin{equation}
\mathcal{L}(A,\gamma) = \prod_{\rm j>i} \mathcal{L}_{\rm ij} \, ,
\end{equation}
where for each $ij$ pair of observations, an underlying Gaussian distribution of $\Delta m$ values is assumed:
\begin{equation}
\mathcal{L}_{\rm ij} = \frac{1}{\sqrt{2\pi \sigma^2(\Delta m)}} \exp\left( -\frac{\Delta m_{\rm ij}^2}{2 \sigma^2(\Delta m)}\right)\, .
\end{equation}
From the model above, the variability of the object, described by a power law, is naturally introduced
in the definition of the variance  $\sigma(\Delta m)^2$ of the underlying Gaussian distribution  as
 \begin{equation}
\sigma^2(\Delta m) = \left[ A (\Delta t_{ij})^\gamma \right]^2 + (\sigma_i^2 + \sigma_j^2) \, .
\end{equation}
The $A$ and $\gamma$ parameters were then obtained by maximization of the likelihood $\mathcal{L}(A,\gamma)$ with the {\sc minuit} package.\footnote{http://wwwasdoc.web.cern.ch/wwwasdoc/minuit/min\-main.html}

We found that only the $g$, $r$ and $i$ bands had useful discriminating power:  quasars have little flux in the $u$ band due to the Lyman continuum absorption of the intergalactic medium for rest frame wavelengths below 91.2~nm, and both $u$ and $z$-band light curves exhibit more noise than the other light curves due to observational limitations (imaging depth and sky background variations in the u and z
  bands). 

The
fitted value of the $\gamma$ parameter is roughly independent of the band.  The fitted amplitudes in the different bands are strongly correlated but not identical. For instance, the $g$ band amplitude is on average larger than the $r$ band amplitude by  about 0.04. To reduce the uncertainty on the 
fitted parameters, we therefore chose to fit simultaneously the $g$, $r$ and $i$ bands for a common $\gamma$ and three amplitudes $(A_{\rm g}, A_{\rm r}, A_{\rm i})$.  We observe an excellent correlation between the amplitudes fitted with a common $\gamma$ and those fitted with an independent $\gamma$ per band, which implies that the data are indeed consistent with a unique power law valid for all bands.

\begin{figure}[h]
\begin{center}
\epsfig{figure=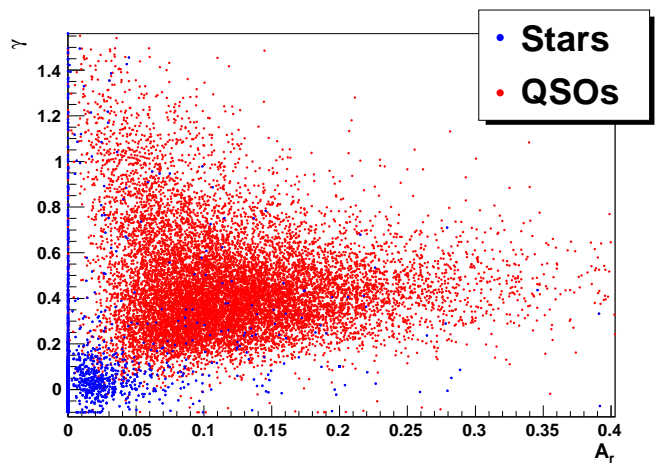,width = \columnwidth} 
\caption[]{Parameters $\gamma$ and $A_{\rm r}$ of the variability structure function for the stellar (blue points) and quasar (red points) test samples. Large $A$'s indicate large fluctuation amplitudes. Large $\gamma$'s indicate an increase of the fluctuation amplitude  with time.} 
\label{fig:AGamma}
\end{center}
\end{figure}

The range of values obtained for stars and quasars are shown in Fig.~\ref{fig:AGamma}. Non variable objects (mostly stars) lie near the origin of the graph, while quasars populate the region of larger $A$ and $\gamma$ values. It is interesting to notice that this approach can also distinguish various variable populations. RR-Lyrae, for instance, can have large variations (thus large $A$) but with no (or little) trend in time, implying that $\gamma$ remains small. The necessary discrimination against variable stars, however, implies that quasars that exhibit a star-like variability cannot be found by this method. The same is even more true for non-variable quasars. 

\subsection{Variability selection of quasars using a Neural Network}\label{sec:NN}

\begin{figure}[htb]
  \centering
 \includegraphics[width=6.5 cm]{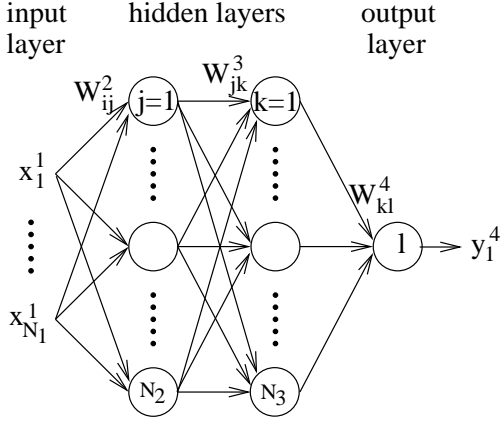}
     \caption{ Schematic representation of the artificial neural network used here with 
$N_1$  input variables, two hidden layers, and one output neuron.}
        \label{fig:ArchitectureNN}
  \end{figure}

To complete our method for discriminating stars from quasars, an artificial Neural Network (NN) was used~\citep{bib:bishop95}.\footnote{We used a C++ package, TMultiLayerPerceptron, developed in the ROOT
environment~\citep{bib:brun95}.} The basic building block of the NN architecture
is a processing element called a neuron.
The NN architecture used in this study is 
illustrated in Fig.~\ref{fig:ArchitectureNN}, where
each neuron is placed on one of four ``layers'', 
with $N_l$ neurons in layer $l$.

The input of each neuron on the first (input) layer is 
one of the $N_1$ variables defining an object. Despite a lesser discriminating power of the $u$ and $z$ bands compared to $g$, $r$ and $i$, the $\chi^2$'s are robust quantities that can be used for all five bands. This is not the case for the structure function parameters, which result from a non-linear fit and were restricted to $gri$. Therefore, for the present study, the chosen variables are the four structure function parameters ($\gamma$, $A_{\rm g}$, $A_{\rm r}$ and $A_{\rm i}$) and the five $\chi^2$'s, leading to $N_1=9$. 

The inputs of neurons on subsequent layers ($l=2,3,4$) are the $N_{l-1}$ outputs (the $x^{l-1}_j ,\,j=1,..,N_{l-1}$) of the previous layer. The inputs of any neuron are li\-near\-ly
combined according to ``weights'' $w^l_{ij}$ and ``offsets'' $\theta^l_j$: 
\begin{equation}
y^l_j=\sum_{i=1}^{N_l} w^l_{ij}\, x^{l-1}_i +  \theta^l_j\,\,
\hspace*{5mm}l\,\geq\,2 \;.
\end{equation}
The output of neuron $j$ on layer $l$ is then defined by the 
non-linear function
\begin{equation}
x^{l}_j = \frac{1}{1+ \exp\left(-y^l_j\right)}\,\,
\hspace*{5mm} 2\leq \,l\,\leq 3 \;.
\label{eq:activation}
\end{equation}
The fourth layer has only one neuron giving an output  $y_{\rm NN}\equiv y^4_1$  reflecting the  strength of quasar-like variability (as probed by the training sample) of the object defined by the $N_1$ input variables.

Certain aspects of the NN procedure, especially the number of layers and the 
number of nodes per layer, are somewhat arbitrary. They are chosen by experience
and for simplicity.
In contrast, the weights and offsets must be optimized so that
the NN output, $y_{\rm NN}$,  correctly reflects the probability
that an input object is a quasar.  To determine the weights and offsets,
the NN must therefore be ``trained''
with a set of objects that are spectroscopically known
to be either quasars or stars. This is done with the test samples described in Sec.~\ref{sec:samples}.

\begin{figure}[h]
\begin{center}
\epsfig{figure=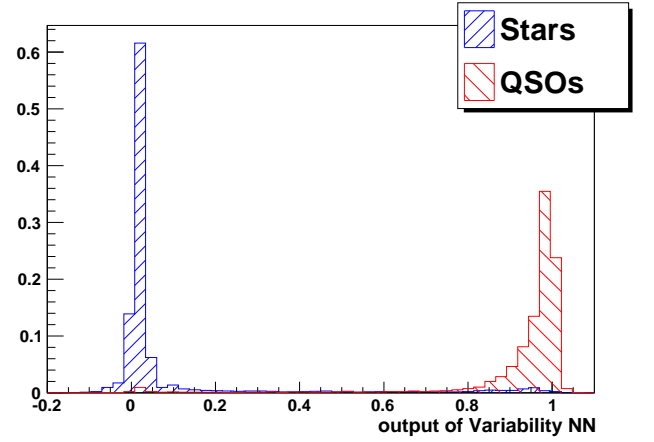,width = \columnwidth} 
\caption[]{Output of the variability Neural Network for the star and quasar samples. 97\% of the quasars have $y_{\rm NN}>0.5$, and 3\% are classified as star-like based on their variability ($y_{\rm NN}<0.5$). The histograms are normalized.} 
\label{fig:NN}
\end{center}
\end{figure}

The result of the NN output is illustrated in Fig.~\ref{fig:NN}. As expected,  most stars peak near 0 while quasars usually have  an output value near 1, and very few objects appear in the middle range where the  variability-based classification is uncertain. 

Only 383 quasars out of 13063 (3\%) are not classified as ``quasar-like'' by the variability NN, i.e. yield $y_{\rm NN}<0.5$. A visual inspection of their light curves confirms that they exhibit no clear variability, neither on short nor
on long time-scales. A minimum loss of $\sim$3\%  is therefore to be expected for any variability-based algorithm to select
quasars using these data. This loss approaches 5\% for the subsample of 3571 quasars at $z>2.15$, probably due to the lower photometric precision
of the objects. Part of the loss might also be due to the smaller rest frame time gap at high redshift. 

\begin{figure}[h]
\begin{center}
\epsfig{figure=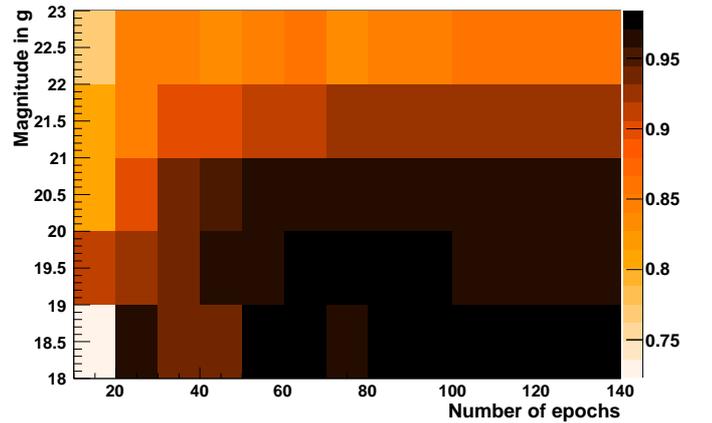,width = \columnwidth} 
\caption[]{$y_{\rm NN}$ (color map) for the quasar sample, as a function of magnitude in g and number of epochs. } 
\label{fig:yNN_mag_NbEpoch}
\end{center}
\end{figure}
$y_{\rm NN}$ is independent of the time span. On average, for quasars,  $y_{\rm NN}$ increases slightly with the number of epochs, as shown in Fig.~\ref{fig:yNN_mag_NbEpoch},  reaching its asymptotic value for about 40 epochs. It also depends on the object magnitude, with a shift of about 0.1 on average between $g\simeq 22.5$ and $g\simeq 18.5$. Most of the objects in Stripe 82 are well-sampled and bright enough not to be affected by these small variations of performance. The results given hereafter are obtained after integration over the full distributions in magnitude and in number of epochs of the quasar sample. 

To quantify the performance of our quasar selection, we define the completeness $C$ and the purity  $P$:
\begin{eqnarray}
C &=& \frac{\rm Number\; of \;selected \;quasars}{\rm Total \;number \;of \;confirmed \;quasars} \;, \label{eq:C}\\ 
P &=& \frac{\rm Number\; of \;selected \;quasars}{\rm Total \;number \;of \;selected \;objects} \;.
\end{eqnarray}
We also define the stellar rejection $R$ as
\begin{equation}
R = 1 - \frac{ \rm Number\; of \;selected \;stars}{\rm Total \;number \;of \; stars\; in\;the\;  sample}\: . 
\end{equation}

\begin{figure}[h]
\begin{center}
\epsfig{figure=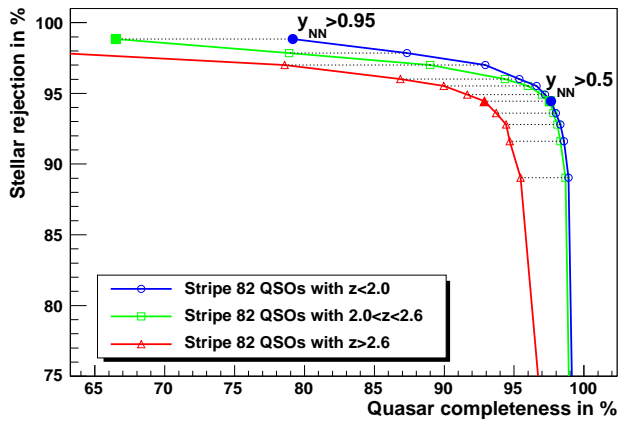,width = \columnwidth} 
\caption[]{Stellar rejection $R$ vs. quasar completeness for the variability-based NN. Open circles are for known quasars at redshift $z$ below 2.0, squares for those with $2.0<z<2.6$  and triangles for those with $z>2.6$. Filled symbols at $R\sim 94.5\%$ and $R\sim 99\%$ indicate the location, on these curves, of the selection thresholds used in Sec.~\ref{sec:bonussel} and \ref{sec:ancillarysel}. } 
\label{fig:performance}
\end{center}
\end{figure}

Fig.~\ref{fig:performance} illustrates the performance, in terms of quasar completeness and stellar rejection, of the variability-based NN,  splitting the quasar sample in three different redshift ranges. For an identical stellar rejection, the loss of quasar completeness with increasing $y_{\rm NN}$ is enhanced at high redshift. 

\begin{figure}[h]
\begin{center}
\epsfig{figure=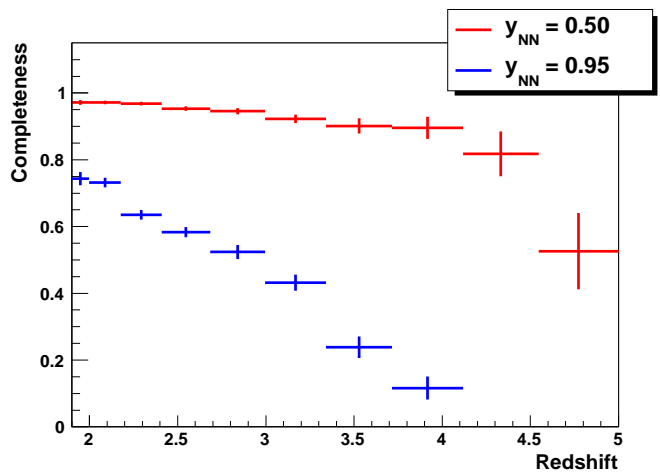,width = \columnwidth} 
\caption[]{Completeness $C$ vs. redshift for two thresholds on the output of the variability NN corresponding to those used for the selections of Sec.~\ref{sec:bonussel} (main sample, with $y_{\rm NN}>0.50$) and \ref{sec:ancillarysel} (extreme variability sample, with $y_{\rm NN}>0.95$).}
\label{fig:effvsz}
\end{center}
\end{figure}

The small redshift-dependence of the variability-based selection method is further confirmed in Fig.~\ref{fig:effvsz}, which  shows the completeness 
$C(z)$ for the two thresholds on $y_{\rm NN} $ used in Sections~\ref{sec:bonussel} and \ref{sec:ancillarysel}. In contrast to a standard quasar selection based on colors, the completeness obtained here 
depends monotonously on redshift and has no minimum at any particular redshift. For a loose cut on the output of the variability NN ($y_{\rm NN}>0.50$), a high completeness is achieved at
all redshifts.
As the cut is tightened ($y_{\rm NN}>0.95$), however, a strong decrease with redshift appears, due to the reduced elapsed  rest-frame time at high redshift, and to
the decrease in the light curve signal-to-noise ratio as objects become fainter, resulting in a weaker significance 
of the variability. 
Nevertheless, even with a tight cut, the method still does not introduce any sharp redshift-specific
feature.

The purity of the selection cannot be determined as easily since it refers to a reference sample.  The training sets are subsamples of the target population (they do not include, for instance, quasars selected through their variability but not through their colors). Knowledge of the total number of selected objects requires a complete sample of targets. Purity will therefore be given in Sec.~\ref{sec:results}, for two cases where the variability selection has been applied to actual data. 

\section{Variability-based selection on Stripe 82 for BOSS}\label{sec:targetting}
BOSS is aiming at a density of $\sim 20\,{\rm deg}^{-2}$ quasars at redshifts $z>2.15$ (hereafter called ``high-$z$'' quasars), with an allocation of $40\,{\rm deg}^{-2}$ optical fibers to obtain spectra of quasar candidates. In this context, the above study can be applied with two major goals. 

The first one is to improve significantly the purity of the list of quasar candidates for which the spectra will be obtained. In BOSS, a traditional color-based selection with single epoch photometry typically reaches a quasar density of 10--15~${\rm deg^{-2}}$ from an initial selection of $\sim 40\,{\rm deg^{-2}}$ targets.
An algorithm with a higher purity presents the advantage of reaching the desired quasar density for BOSS, meaning an increase of about a factor 2, while keeping the number of fibers fixed. This is the aim of the ``Main sample'' described in Sec~\ref{sec:bonussel}.  

The second goal is to search for additional quasars, that would have been missed by previous searches because of colors beyond the typical range considered so far for quasars, but that could be selected based on their variability. This is the strategy leading to the selection of the ``Extreme variability sample'' presented in Sec.~\ref{sec:ancillarysel}. These targets are expected to constitute a sample that would be less biased with redshift than through color selections. It would contribute to improving our knowledge of the quasar population in the  approximate redshift range  between 2 and 4.

Both approaches were adopted by BOSS for the  observation of Stripe 82 in September and October 2010. The results obtained are given in Sec.~\ref{sec:results}, and a comparison with color-based selections is presented in Sec.~\ref{sec:discussion}.  

\subsection{Main sample}\label{sec:bonussel}
The goal of the Main sample was to obtain a list of about 35~$\rm deg^{-2}$ targets with high quasar purity.

\begin{figure}[h]
\begin{center}
\epsfig{figure=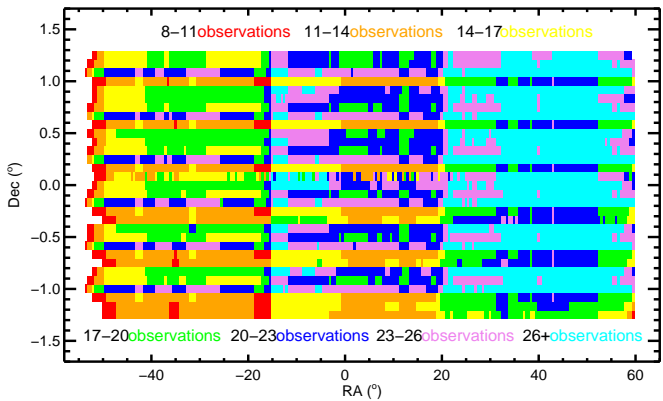,width = \columnwidth} 
\caption[]{Number of SDSS-I and SDSS-II measurements used to derive the co-added photometry in Stripe 82.} 
\label{fig:coadd}
\end{center}
\end{figure}

A color-based analysis with very loose thresholds is used to yield an initial list of $\sim70 \,{\rm deg}^{-2}$ objects, expected to be dominated by stars by at least a ratio 2:1.  Quasars are seen to have varying colors with time, since their structure function amplitudes $A$ are band-dependent while the power $\gamma$ is unique for all bands. However, the color change over a decade is observed to be small, with an average shift of 0.1 mag only. We thus co-added single epoch observations (cf. Fig.~\ref{fig:coadd}) to improve the photometry of the objects and their color measurements. The criteria for the preselection were defined as follows:
\begin{itemize}
\item output of a color-based NN $> 0.2$ (with colors determined from co-added observations) to remove objects that were far from the quasar locus in color-space~\citep{bib:yeche10},
\item$(u-g)>0.15$ to enhance the fraction of $z>2.15$ quasars over low-$z$ ones. This cut rejects only 1\% of previously known high-$z$ quasars.
\end{itemize}
The completeness of this preselection for high-$z$ quasars is of order 85\%, which corresponds to an upper bound on the completeness of the ``main sample''. 

Requiring $y_{\rm NN}> 0.50$, and removing previously identified low redshift quasars, we obtained a selection of 7586 objects (i.e. a target density of $34.5 \,{\rm deg}^{-2}$), called hereafter the ``main sample". Technical reasons related to the tiling of the objects~\citep{bib:blanton03} reduced this sample to a density of $31.1\,{\rm deg}^{-2}$. As shown in Fig.~\ref{fig:effvsz}, the completeness of the variability selection at this threshold is expected to be $\sim 95\%$ (of the sample to which it is applied). 

 For comparison with the more usual color selection, we can remove the final variability selection and replace it by a tightened color cut (still using co-added photometry) adjusted to also produce a sample of 7586 targets. This color-selected sample and the main sample  have 73\% of their targets in common. As clearly visible in Fig.~\ref{fig:NN}, the threshold of 0.50 is very loose. There is thus no additional gain to be expected by lowering further the variability threshold. 

\begin{figure}[h]
\begin{center}
\epsfig{figure=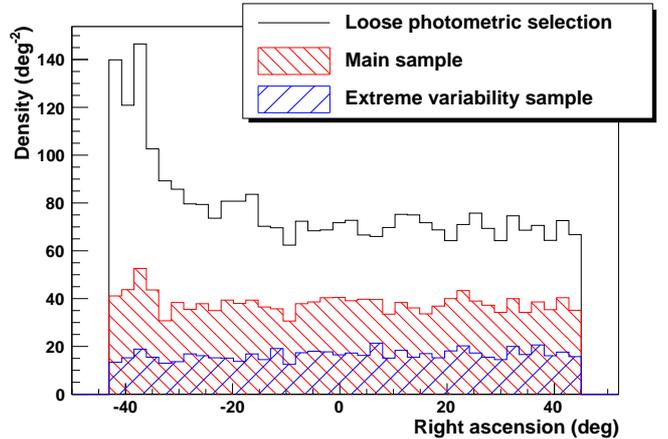,width = \columnwidth} 
\caption[]{Right Ascension distribution of targets in the main sample at the stage of loose color-based selection (black histogram), and after the final variability-based selection (red histogram). The targets of the extreme variability program are shown as the blue histogram. } 
\label{fig:rabonus}
\end{center}
\end{figure}

Fig.~\ref{fig:rabonus} shows that the target density is flat with Right Ascension, as expected for extragalactic objects, in contrast to the peak that would be expected for  $\alpha_{\rm J2000}\simeq-43^\circ$ in the case of  large contamination by Galactic stars as is seen in the initial distribution corresponding to the loose photometric preselection. 

\begin{figure}[h]
\begin{center}
\epsfig{figure=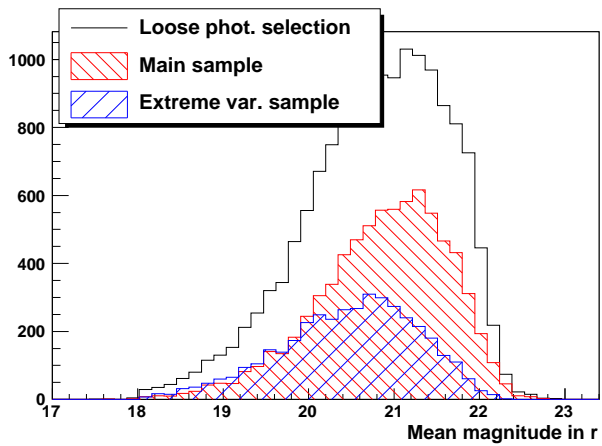,width = \columnwidth} 
\caption[]{ Distribution of the magnitude in $r$  at the stage of loose color-based preselection (black histogram), and after the final variability-based selection leading to the main sample (red histogram). The targets of the extreme variability program are shown as the blue histogram. }
\label{fig:mr_distrib}
\end{center}
\end{figure}

 Fig.~\ref{fig:mr_distrib} shows the distribution of the magnitude in the $r$ band for the different samples. The drop at $r>21$ is due to the color preselection. The selection leading to the main sample (red histogram) does not change the shape of the initial distribution (black histogram). This agrees with the fact that little redshift (and magnitude) dependence is observed at a threshold of 0.50 on the variability NN (cf. Fig.~\ref{fig:effvsz}). The relative efficiency of the variability selection with respect to the preselected sample is roughly independent of magnitude.

\subsection{Extreme variability sample}\label{sec:ancillarysel}
The second goal was to obtain an independent and complementary list of about $3\, \rm deg^{-2}$ objects selected by the variability NN but rejected according to their colors. With this approach, we could expect to find quasars in the stellar locus, at the risk of obtaining a sample dominated by variable stars rather than by quasars.  This sample, however, offers a unique opportunity to explore a new region of color-space. Given the high level of discrimination between quasars and stars that is seen Figs.~\ref{fig:AGamma} and \ref{fig:NN}, the extreme variability sample is expected to have a strong potential.
 
The total number of point-like objects in Stripe 82 is on the order of several millions. Because the computation of the variability parameters on such a large sample would have been both disk- and time-consuming, a very loose preselection of about $1000\,\rm deg^{-2}$ objects was first applied, with the following criteria:
\begin{itemize} 
\item $i>18$ to limit the contribution from low-$z$ quasars but $g<22.3$ to maintain the possibility to obtain a good spectrum, 
\item $(g-i)<2.2$ to exclude M stars,
\item $(u-g)>0.4$ to enhance the fraction of $z>2.15$ quasars compared to low-$z$ ones,
\item $c_1<1.5$ or $c_3<0$ to remove a region in color-space distant from quasars and strongly populated by stars, where colors $c_1$ and $c_3$ are defined in~\cite{bib:Fan} as 
\begin{eqnarray}
c_1 &=& 0.95(u-g)+0.31(g-r)+0.11(r-i) \, ,\nonumber \\
c_3 &=& -0.39(u-g)+0.79(g-r)+0.47(r-i)\, .\nonumber 
\end{eqnarray} 
\end{itemize}

While these cuts reduced the total number of objects by about a factor of ten, leading to a sample of about 235,000 targets over the 220 ${\rm deg}^{-2}$ area of Stripe 82, they rejected only about 9\% of previously known quasars at $z>2.15$,  uniformly over the magnitude range. 

Requiring $y_{\rm NN}> 0.95$ (i.e. selecting the most variable objects)  then yielded  a sample of 4360 targets (or a density of $\sim 20\,{\rm deg}^{-2}$)  called hereafter the ``extreme variability sample". Not all the targets could be observed: technical limitations  (allocated number of fibers and tiling) reduced this sample to a density of $\sim 15 \,{\rm deg}^{-2}$.

The distribution of the Right Ascension of the selected objects is shown in Fig.~\ref{fig:rabonus} as the blue histogram. Its flatness is again an indication of low stellar contamination.

The magnitude distribution of this sample is illustrated in Figure~\ref{fig:mr_distrib} as the blue histogram: the selection efficiency drops by about a factor of two between the maximum, for a magnitude near  20, and its level at magnitudes near 22. This drop is to be expected given the decrease of completeness with redshift shown in Fig.~\ref{fig:effvsz}.

 About 65\% of the  extreme variability-selected quasars is also part of the main sample of Sec.~\ref{sec:bonussel}. Because of the technical limitations mentioned above, which are tighter for the extreme variability sample than for the main one, the overlap increases to  78\% of the actual targets. The remaining targets constitute what we call hereafter  the  ``extreme variability {\it only} sample".  It contains 748 objects (i.e. a density of $3.4 \,{\rm deg}^{-2}$) for which spectra were measured.  

\subsection{Results}\label{sec:results}

\begin{figure*}[htbp]
\begin{center}
\includegraphics[angle=-90, width = .65\columnwidth] {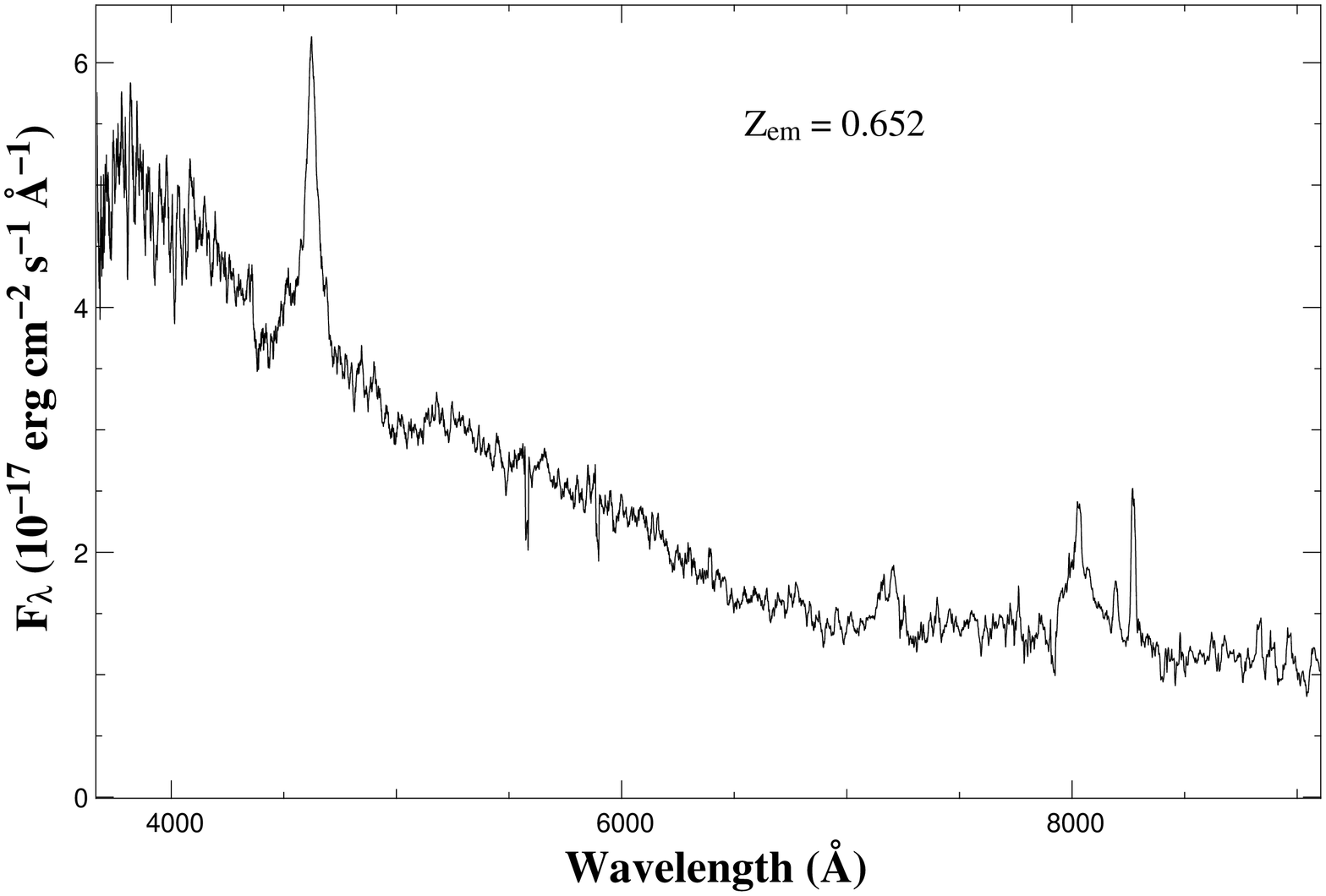} \hspace{.2cm}
\includegraphics[angle=-90, width = .65\columnwidth] {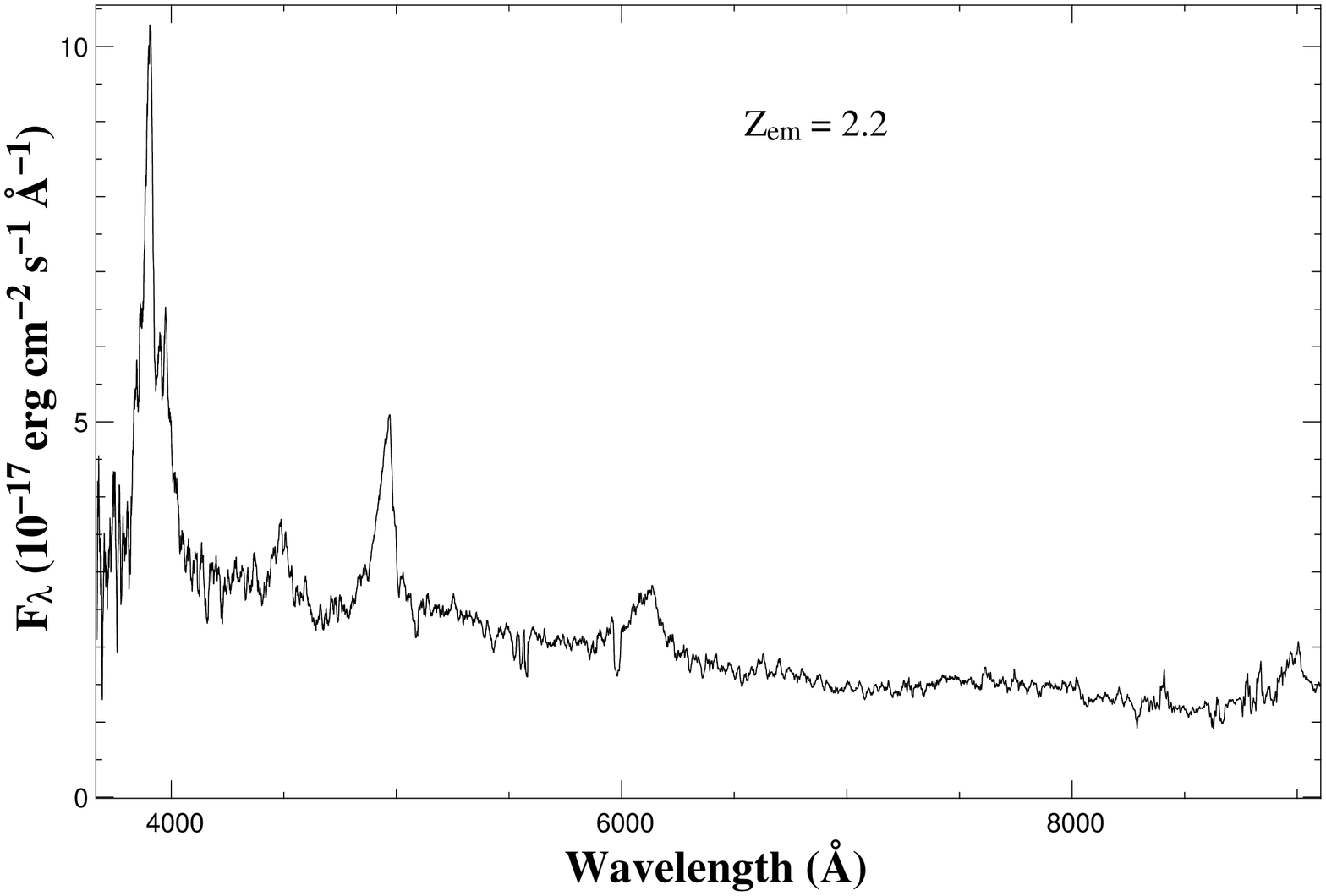} \hspace{.2cm}
\includegraphics[angle=-90, width = .65\columnwidth] {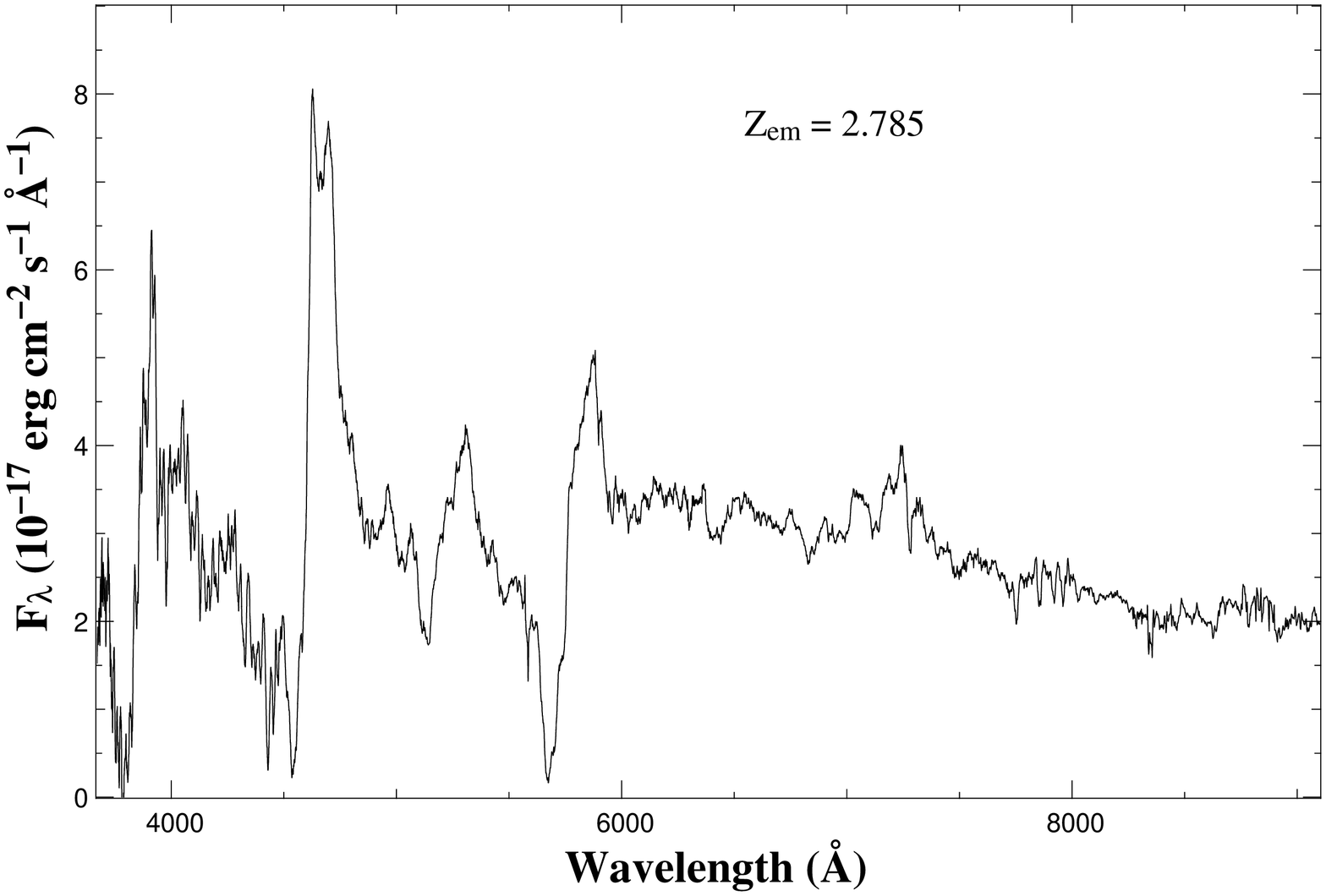}  
\includegraphics[angle=-90, width = .65\columnwidth] {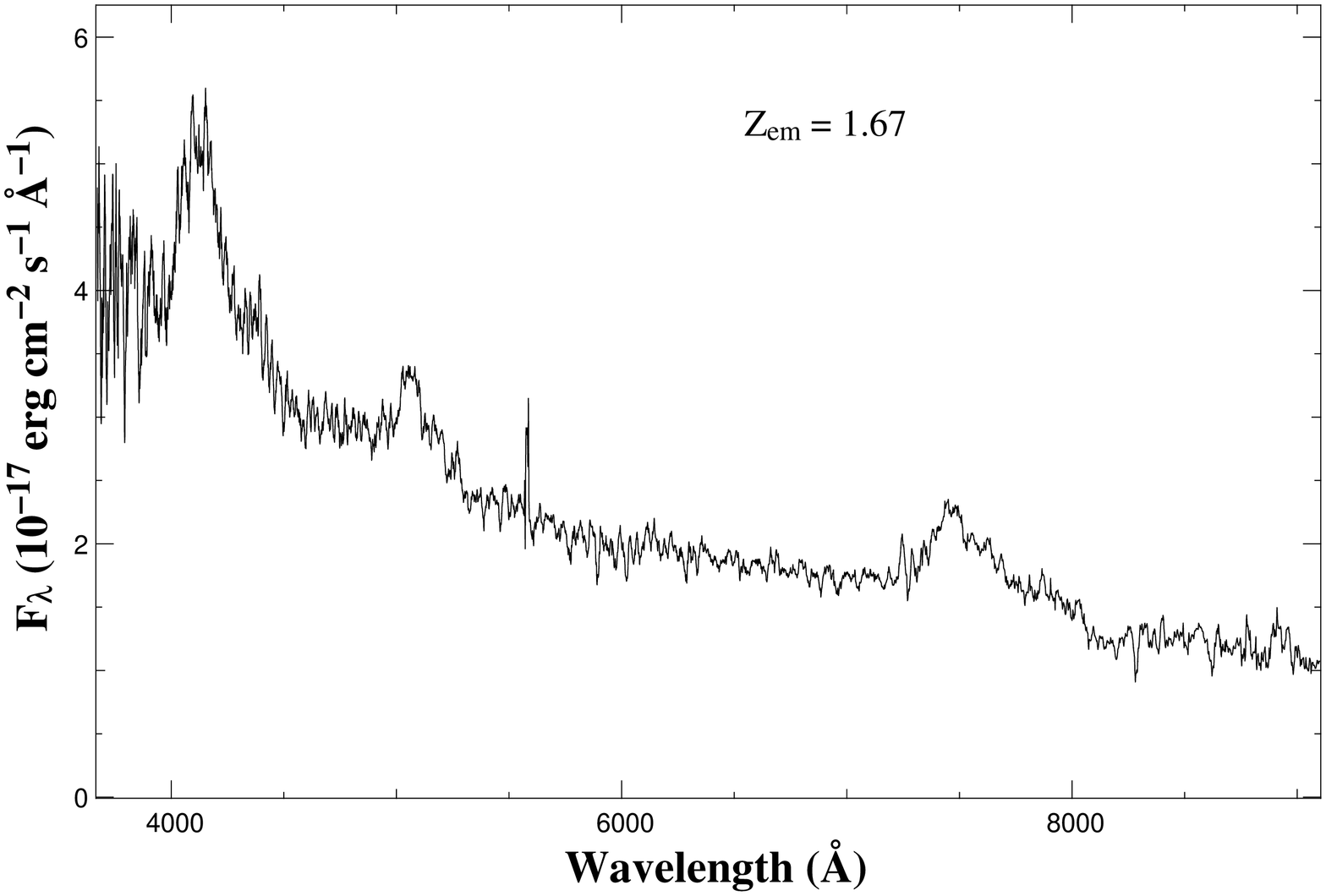}  \hspace{.2cm}
\includegraphics[angle=-90, width = .65\columnwidth] {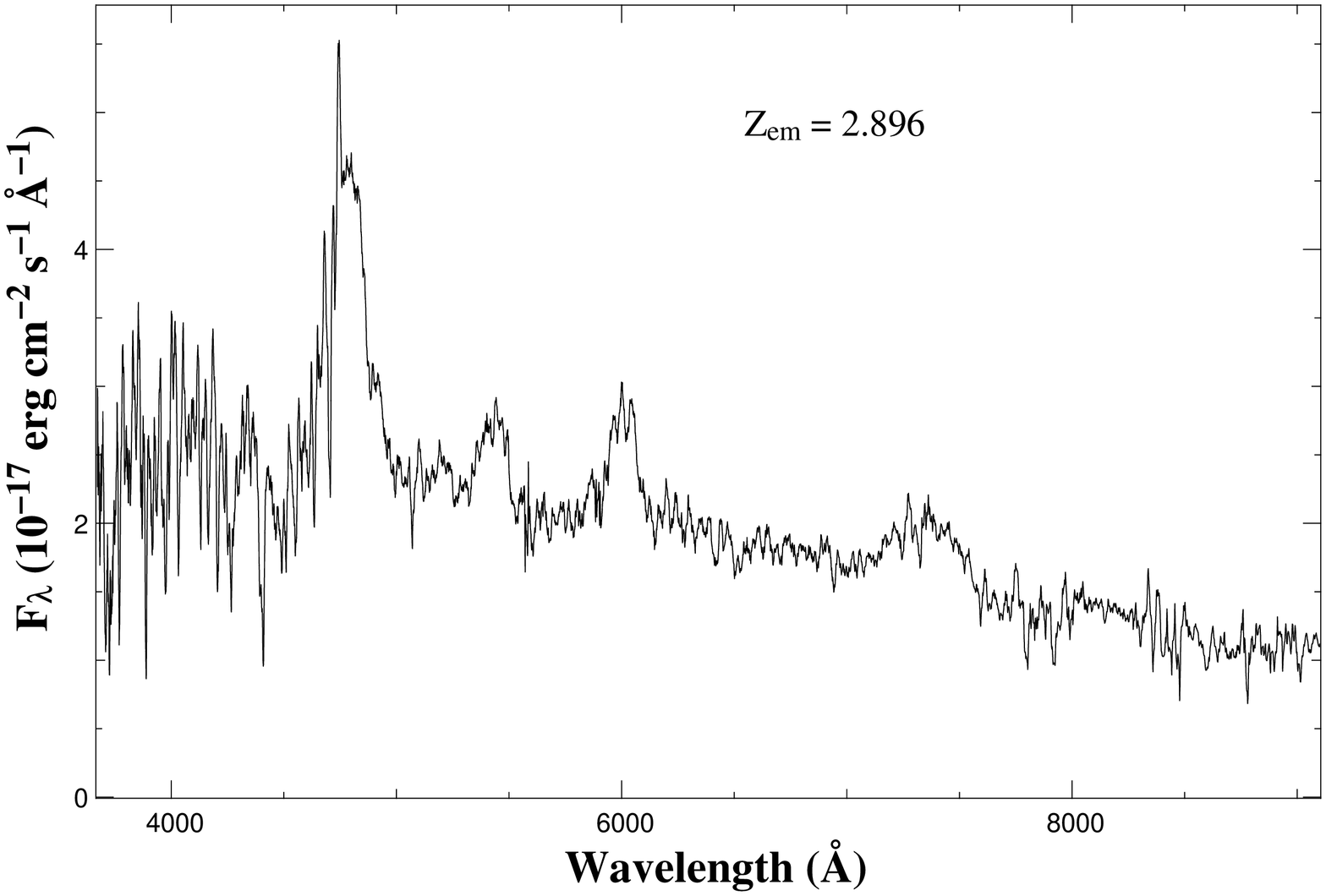}  \hspace{.2cm}
\includegraphics[angle=-90, width = .65\columnwidth] {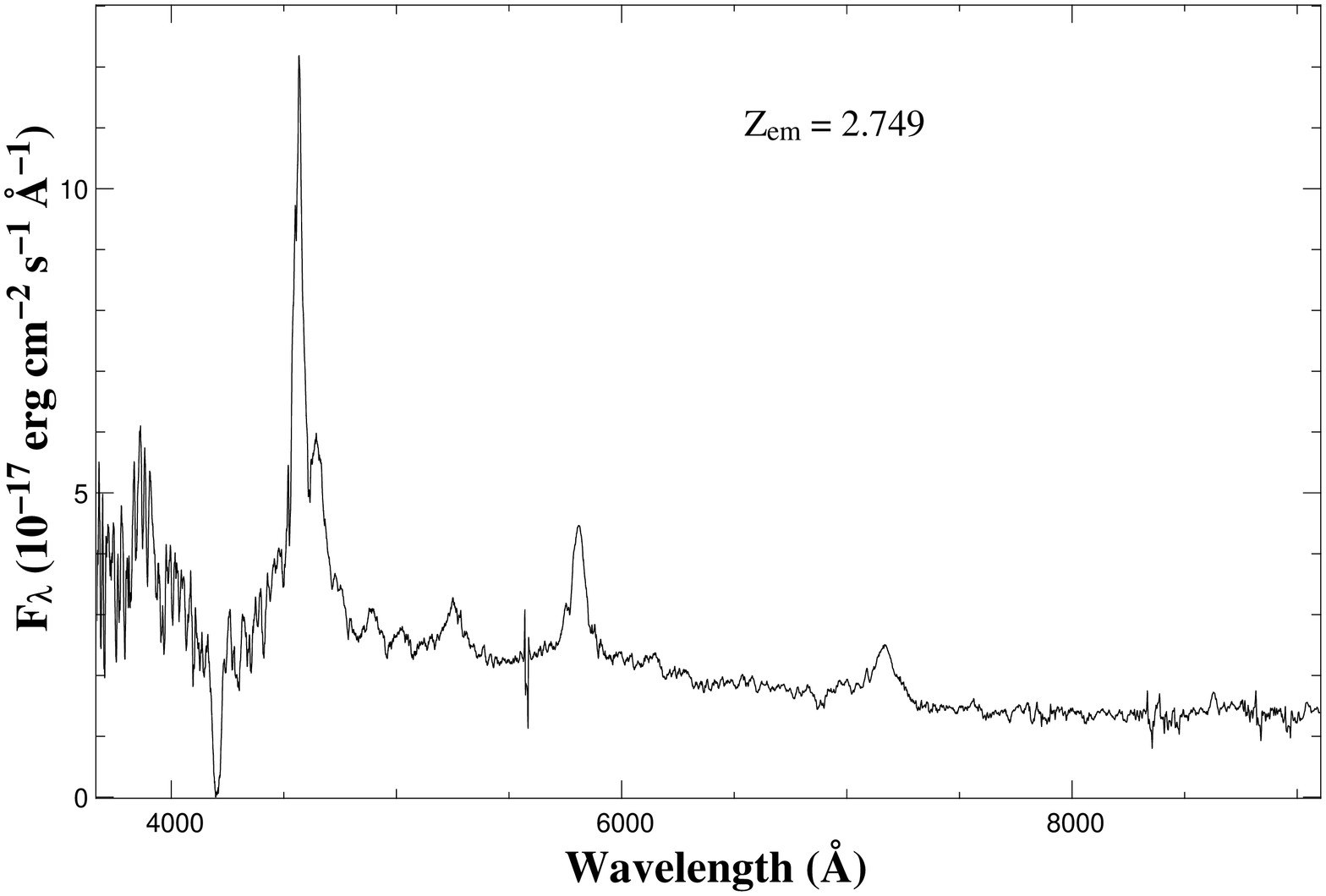} 
\caption[]{ Selection of quasar spectra from the variability targets, here shown smoothed over 9~\AA. Upper and lower left: low-$z$ quasars. Upper and lower middle: high-$z$ quasars. Upper right: Broad Absorption Line high-$z$ quasar. Lower right: high-$z$ quasar displaying a Damped Lyman-$\alpha$ absorption. } 
\label{fig:spectra}
\end{center}
\end{figure*}

\begin{table*}[htb]
\begin{center}
\begin{tabular}{lc|cc|ccc|ccc|ccc}
 \hline
 Selection & Target & \multicolumn{2}{c|}{All quasars} &  \multicolumn{3}{c|}{$z>2.15$}&  \multicolumn{3}{c|}{$2.15<z<3.0$}&  \multicolumn{3}{c}{$z>3.0$}  \\ 
 sample& density & Density & $P$(\%)  & Dens.& $P$(\%)  & $C$(\%)& Dens.& $P$(\%)  & $C$(\%) & Dens.& $P$(\%)  & $C$(\%)  \\
 \hline
Main sample & 31.1 & 29.0 & 93 & 22.3& 72&84& 18.1& 58& 86 & 4.2 &14& 76  \\
Extreme var.  &  15.1 & 14.6 & 96 & 12.1& 80 & 45& 10.4 &68 &49 & 1.7 &11 &31\\
Extreme var. only  &  3.4 & 2.9 & 86 & 1.7 & 49 & 6& 1.4 &41 &7  & 0.3 &8 &5\\\hline
Total & 34.5 & 31.9 & 92 & 24.0& 69& 90& 19.5 &56& 92 & 4.5 &13 &81 \\
\hline
\end{tabular}
\caption[]{Density, purity $P$ and completeness $C$ of variability-based selections of quasar candidates. Densities are in ${\rm deg}^{-2}$ over an area of $220\,{\rm deg^{-2}}$. Purity is the ratio of the density of the quasars in a given sample to the target density. Completeness includes all identified high-redshift quasars, whether from their color, variability, radio emission, etc. Column ``Target'' is for all candidates, ``All quasar'' refers to confirmed quasars independently of their redshift. Line ``Extreme var." includes both the extreme variability sample and the main sample targets that fulfilled the requirement $y_{\rm NN}>0.95$. Line ``Extreme var. only'' refers to objects rejected from the main sample due to their colors.   }
\label{tab:results}
\end{center}
\end{table*}

 Thanks to good weather conditions, all planned targets have been observed. The reduction of the spectra was performed by the BOSS pipeline~\citep{bib:bolton}, which also gives a preliminary determination of the redshift of the identified quasars. All spectra were then checked visually to yield final identifications and redshifts.  Special features such as Broad Absorption Line (BAL) quasars were identified during this visual inspection. The pipeline and visual scanning are in agreement for $\sim 95\%$ of the objects.  The spectra will be made available with the SDSS data release DR9, expected for mid-2012. A small selection is given in Fig.~\ref{fig:spectra}.

The outcome of the targeting of the two samples described above is summarized in Table~\ref{tab:results}.  A total of 5270 high-redshift quasars were confirmed (4900 in the main sample, 2650 in the extreme variability sample of which 370 not in common with the main sample), a 
significant improvement over previous results. About half of these quasars (2770) were not known previously and were revealed by the present study. As stated in the abstract, we see that 90\% of the known high-redshift quasar population is recovered by its variability, and that 92\% of the selected targets are quasars (i.e.,  only 8\% non-quasars). This high purity is in
agreement with the flat Right Ascension distributions of the two samples shown in Fig.~\ref{fig:rabonus}, indicating negligible stellar contamination.  

The main sample has a quasar purity of 93\% on average and 72\% at a redshift $z>2.15$.
From this sample alone, the average density of $z>2.15$ quasars over Stripe 82 has been increased from $\sim 15\,{\rm deg}^{-2}$ from previous BOSS observations to $22.3\,{\rm deg}^{-2}$.

It is remarkable that  86\% of the objects in the ``Extreme var. only'' category, all rejected according to their colors, are quasars. Half of these, furthermore, are at $z>2.15$. These results confirm the expected potential of the extreme variability program.

Considering the full sample selected from its extreme variability (i.e. including the candidates in the main sample that fulfilled the requirement $y_{\rm NN}>0.95$, cf. line ``Extreme var." of Table~\ref{tab:results}), we achieve an even higher purity: 96\% of the objects are quasars, and 80\% are at a redshift above 2.15. These results imply that variability is indeed an efficient tool for selecting quasars against all other variable sources.

 The results for high-redshift quasars are also given split into two redshift bins. The drop of completeness with redshift expected from Fig.~\ref{fig:effvsz} for the extreme variability sample appears clearly. This sample, much more than the main sample does,   selects preferentially quasars in the $2.15<z<3.0$ than in the $z>3.0$ bin: the respective purities in the two bins are 68\% and 11\% for the extreme variability sample vs. 58\% and 14\% for the main sample.

The low fiber budget allocated to the Extreme variability  program does not make the study of its completeness a relevant issue. However, we note that with a target density of only 3~$\rm deg^{-2}$, the extreme variability program  raised the high-$z$ completeness of the main sample by  $\sim$6\%. 

\begin{figure}[h]
\begin{center}
\epsfig{figure=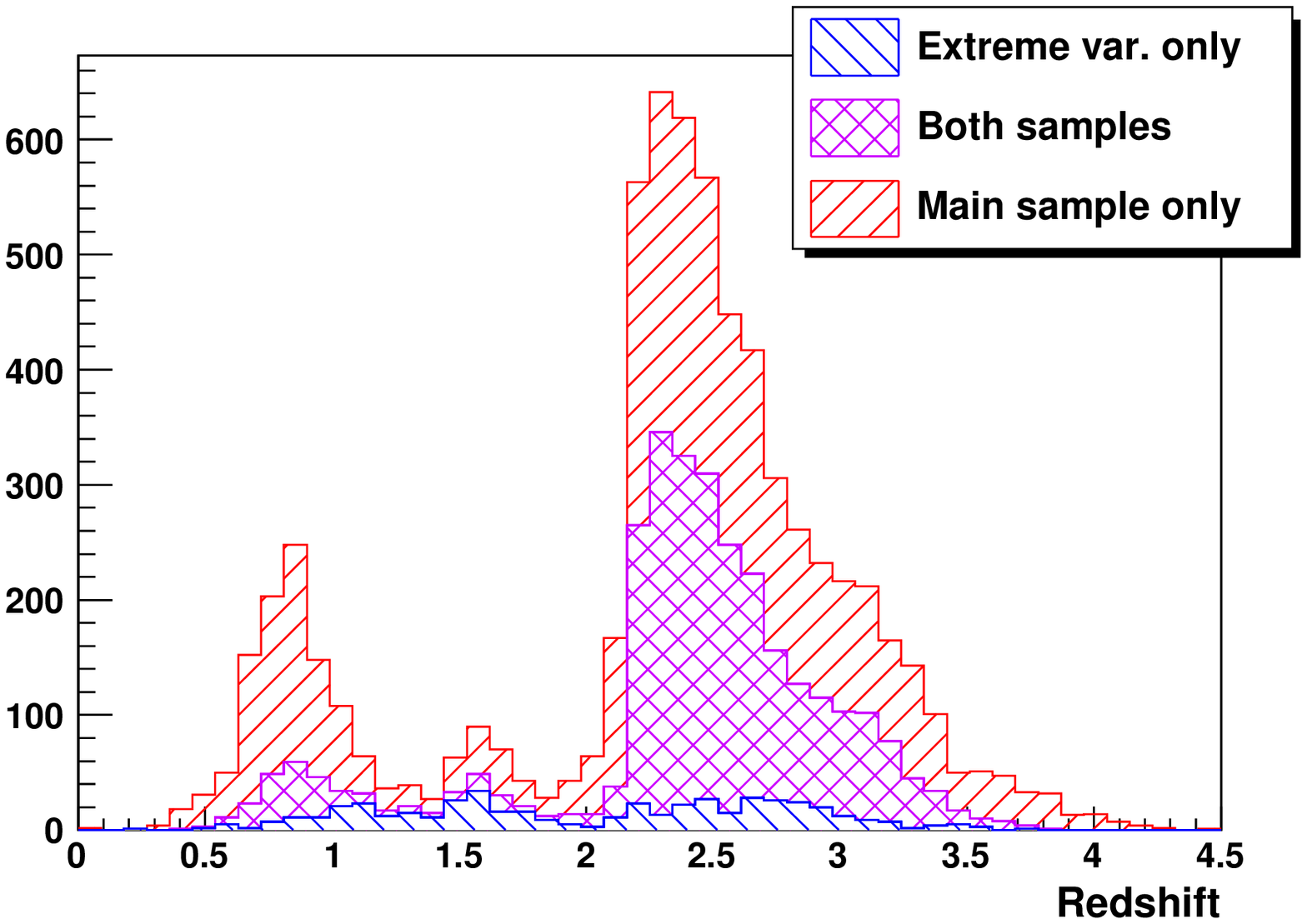,width = \columnwidth} 
\caption[]{Stacked redshift distribution of the confirmed quasars, where the histograms represent the number of quasars in each of the non-overlapping samples. The total extreme-variability sample is thus illustrated by the blue+purple surface, while the total main sample is in purple+red. The emphasis of the selection on $z>2.15$ objects is apparent.} 
\label{fig:z_distrib}
\epsfig{figure=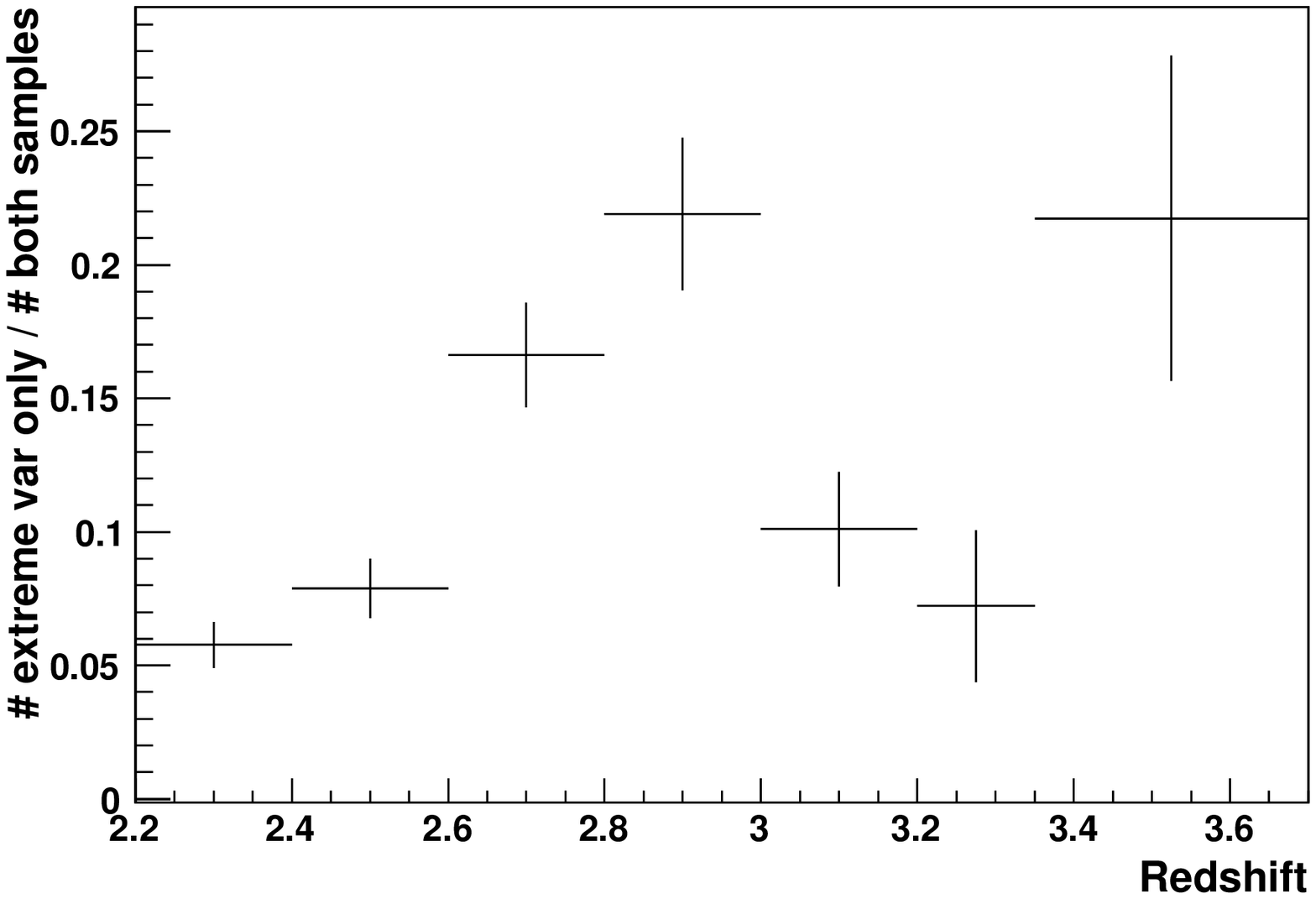,width = \columnwidth} 
\caption[]{Redshift distribution of the fraction of quasars added by the extreme variability selection compared to quasars in the same variability range but fulfilling color constraints. } 
\label{fig:z_apport}
\end{center}
\end{figure}
Fig.~\ref{fig:z_distrib} shows the redshift distribution of the quasar samples selected through variability. 
As expected from the cut on $u-g$,
most are at $z>2.15$, corresponding to the requirements of BOSS. 
Fig.~\ref{fig:z_apport} shows that the additional quasars selected via extreme variability tend to preferentially 
lie in the $2.5<z<3.0$ redshift range where  color-based selections are known to be incomplete. This indicates 
that a pure variability-based selection can indeed contribute to the recovery of quasars lost during the color selection.  
The low number of quasars at $z>3.4$ prevents  firm conclusions from being drawn on this higher redshift range.

\begin{figure*}[ht]
\begin{center}
\epsfig{figure=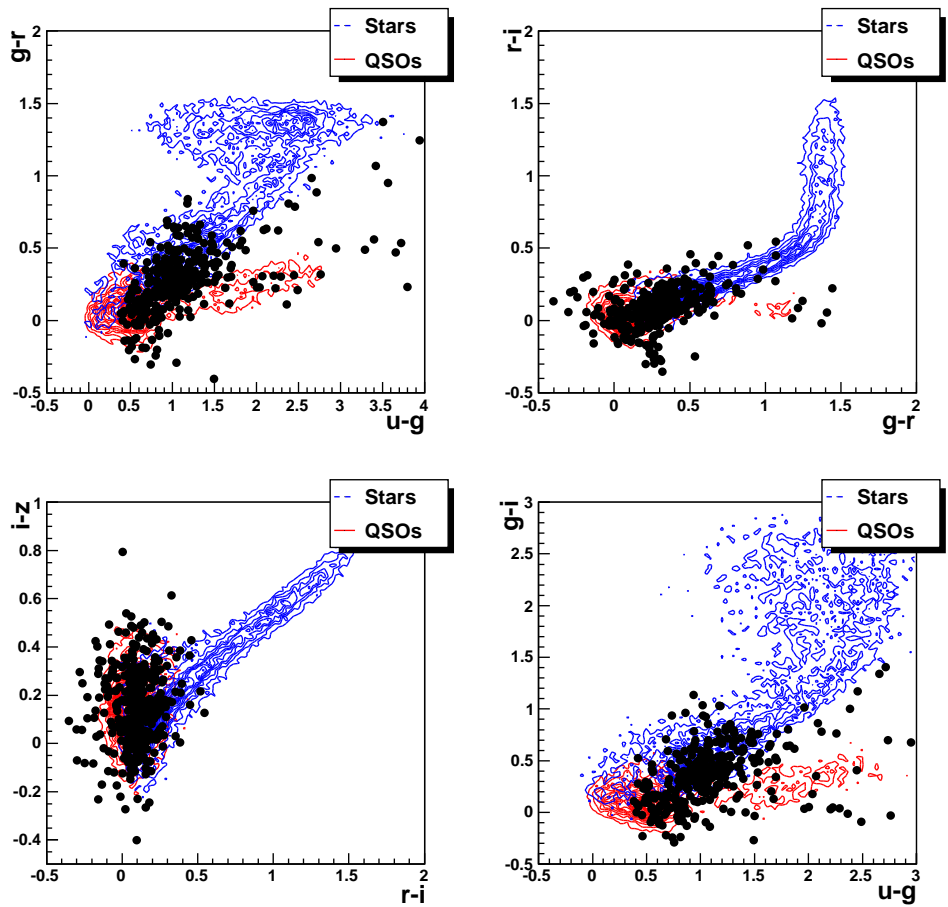,width = 1.5\columnwidth} 
\caption[]{Color-color plots indicating the stellar (blue) and quasar (red) loci, as well as the position of the 370 high-redshift additional quasars
 rejected from their colors but selected through the variability neural network (extreme variability sample described in Sec.~\ref{sec:ancillarysel}).} 
\label{fig:color}
\end{center}
\end{figure*}

The location of the additional quasars in color-color space is presented in Fig.~\ref{fig:color}. There is no indication that they form a new class of quasars; instead, they appear to extend the quasar locus into the stellar locus in all color-color diagrams, as expected from synthetic models of quasar evolution~\citep{bib:Fan}.  The completeness of the extreme-variability sample is quite low (cf. Table~\ref{tab:results}), so we can expect many more quasars than found here to be located in disfavored regions of color-space. High-$z$ quasars are therefore probably even less well separated from the stellar locus than previously thought. 

The fraction of Broad Absorption Line (BAL) quasars among the $z>2.15$ quasars is seen to be higher in the sample selected for its extreme variability than in the main sample that includes  stricter color cuts. Comparing the two non-overlapping ``main" and ``extreme var only" samples, we have
\begin{eqnarray}
\frac{{\rm Number \; of \;  high}\; z \; {\rm BAL \; quasars}}{{\rm Number\;  of \; high}\;  z\;  {\rm quasars}} =&\nonumber \\
 7.0\%\pm 0.4\%&& (\rm Main\; sample) \nonumber \\
14.6\%\pm 1.8\%&& (\rm Extreme\; var. \; only) \nonumber
\end{eqnarray}
This seems to indicate that quasars affected by BAL features tend to fall outside the color regions that are generally  favored by quasars. 

\subsection{Comparison with color selection}\label{sec:discussion}

 We compare the results obtained from this work to color selections of quasars. Two cases are studied below. The first one is 
a traditional  color selection  using single-epoch photometry. The large number of observations in Stripe 82, however, also permits a second approach using photometry obtained on co-added images, i.e. deeper frames and with a higher signal-to-noise, as was used for the color preselection of the main sample.  A color selection on co-added images is expected to  be much more complete than one based on single epoch observations.

\begin{figure}[ht]
\begin{center}
\epsfig{figure=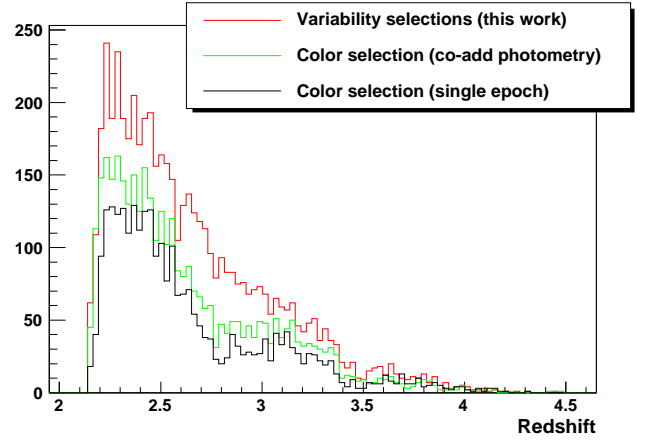,width =\columnwidth} 
\caption[]{Redshift distribution of the quasars recovered for three different selection algorithms presented in the text. } 
\label{fig:z_multiselection}
\end{center}
\end{figure}

In both cases, we  derived lists of 34.5~deg$^{-2}$ targets as for the total variability-based selection (main and extreme variability samples) presented in this paper. We compared the outcome of  these color-based selections to that of the variability-based one, using the full set of quasars identified on Stripe 82 from their color, variability or radio emission. The outcomes of the different selections are in the ratio 0.5:0.7:1 for the single-epoch color selection, co-added color selection and variability (this work) selection respectively.  
Fig.~\ref{fig:z_multiselection} shows the redshift distribution of the quasars recovered from the different samples. The dip around for $2.5<z<3.2$ in both color selections is clearly visible.

The advantage of variability might have been larger still with a greater ratio of the 34.5~deg$^{-2}$ fibers allocated to the extreme-variability sample,  since the latter has a higher purity than the main sample (cf. Table~\ref{tab:results}). As variability and colors seem to yield complementary samples (some quasars can be selected one way and not in the other), the most promising method would be to use both pieces of information simultaneously.

\section{Use of external data and application to the full SDSS sky}\label{sec:PTF_PS1}
Given the success of the variability-based selection in Stripe 82, it would be interesting to apply it over
a much wider area in the sky. One possibility would be to use jointly data from SDSS (one or two photometric measurements
over 10,000 $\rm deg^2$) and forthcoming data from the Palomar Transient Factory (PTF) or Pan-STARRS 1 (PS1),
which cover the same $10,000\,\rm deg^2$ at several occasions over 3 to 5 years. A strategy based on these various data sets can be useful to future surveys  like BigBOSS\footnote{http://bigboss.lbl.gov} or  LSST~\citep{bib:LSST, bib:ivezic}.

\subsection{Extrapolation to PTF}

Since December 2008, PTF has taken data in the $R$ band at the cadence of one measurement every 5 nights~\citep{bib:PTF}. The images can be co-added to produce 4 deep frames per year of observation.
Apart from Stripe 82, most of the area covered by SDSS  was observed only once. The data available for quasar searches at the end of the PTF survey can therefore be expected to consist typically of 1 point from SDSS (useful to extend the lever arm in time lag) and 4 points per year from PTF. 
To explore the possibilities offered by this data combination for quasar selection, we constructed synthetic light curves by down-sampling data from Stripe 82 in the following way:  \\
- The last 5 years of SDSS are used to simulate PTF measurements: four evenly spaced points per year are selected from the SDSS data,\\
- To simulate the sole measurement available from SDSS on most of the sky, one point is taken at random over the previous years of SDSS, maintaining a gap of at least 2 years between the SDSS point and the first PTF measurement (to ensure a realistic lever arm). \\
Only synthetic light curves with all 21 measurements (1 for SDSS and 4 for each of the 5 years of PTF) are considered hereafter. With this constraint, we are left with 2248 (83\%) stellar and 11456 (86\%) quasar light curves (out of the initial samples described in section~\ref{sec:samples}).

\begin{figure}[h]
\begin{center}
\epsfig{figure=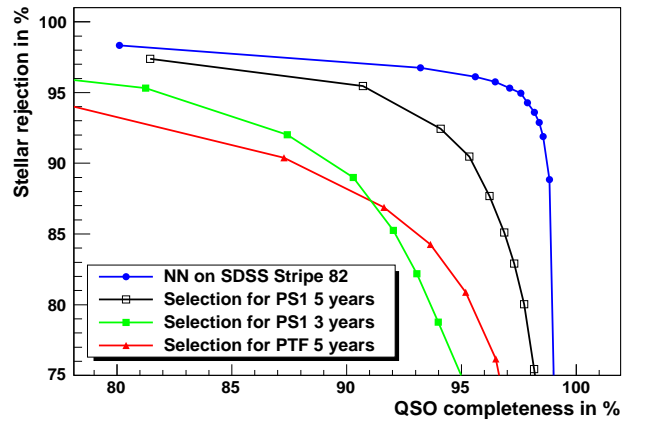,width = \columnwidth} 
\caption[]{Stellar rejection vs. quasar completeness for the full Stripe 82 data (blue dots), for the Pan-STARRS (green and black squares) and for the PTF  (red triangles) simulated data. In each case, the threshold on the relevant variability NN is increased from right to left. } 
\label{fig:PTF}
\end{center}
\end{figure}

As PTF observes only in one band, the variability parameters are reduced to the reduced $\chi^2$ in $r$, $A_{\rm r}$ and $\gamma$. A neural network was trained on the usual stellar and quasar test samples 
to yield an estimator of quasar likelihood based on these 3 parameters. The red triangles in Fig.~\ref{fig:PTF} mark the evolution of the stellar rejection vs. quasar completeness as the threshold on the NN output is varied. They show that one can reach a quasar completeness of 85\% for a rejection of 91\% of the stars.
For comparison, the blue dots illustrate the favorable case of Stripe 82 with all available measurements on 5 bands (case studied in Section~\ref{sec:targetting}) and a variability selection based on the 9-parameter NN. 

Note that as explained in Sec.~\ref{sec:samples}, the stellar sample used for figure~\ref{fig:PTF} has passed loose color cuts that might not be available for PTF data. We have checked that the performance of the algorithm in the rejection of randomly picked  Stripe 82 objects, statistically dominated by stars by at least a ratio 10 to 1, is within 1\% of the performance plotted in the figure. 

\subsection{Extrapolation to PS1}

Pan-STARRS 1 (PS1) started regular observations in March 2009. With its 3 degree field of view, the whole available sky is recorded 
3 times during the dark time of each lunar cycle. The first part of the project is expected to last about 3 years, after which a second telescope will begin operation. To explore the use of the PS1 data, we proceeded in a similar way as for PTF. The main difference is that PS1 has data available in five filters ($g$, $r$, $i$, $z$ and $y$) instead of one. For quasar selection in the redshift range $2.15<z<4$, we considered only the filters in common with SDSS ($g$ through $z$). This restriction produced 8 variability parameters: four $\chi^2$'s (one in each of the four bands), $A_{\rm g}$, $A_{\rm r}$, $A_{\rm i}$ and the common $\gamma$ (as for the study of Stripe 82). As for PTF, a NN was trained to yield an estimator based on these 8 parameters. The performance of the resulting selection is illustrated in Fig.~\ref{fig:PTF} for two survey durations, 3 or 5 years. Only synthetic light curves with all 13 (in the case of a 3-year survey) or 21 (in the case of a 5-year survey) measurements are considered in the plot.

The 3-year survey gives results comparable to those for the 5-year PTF. In contrast, the 5-year PS1 survey is a significant improvement over the 3-year survey, and can reach an 85\% quasar completeness for a 97\% stellar rejection, or a 91\% quasar completeness for a 95\% stellar rejection. 

The absence of the SDSS anchor point would reduce the quasar completeness by about  3\%. Of course, the SDSS data would have little impact on the stellar rejection $R$, since most stars exhibit flat light curves, whatever their coverage. 

\subsection{Extrapolation to fainter high-$z$ targets with PS1}

Quasar selection was typically concentrated at $g<22.3$. Future surveys like BigBOSS intend to go deeper in order to increase the density of quasars. To study the impact of a deeper magnitude limit on the performance of the variability selection,  we used all objects defined as point sources in coadded frames to compute stellar rejection vs. quasar completeness for magnitude limits $g<21$, $g<22$ and $g<23$, in the case of five years of PS1 data. The coadded images are used to detect the sources out to $g>23$, while the lightcurves are still  simulated by downsampling the shallower, single-epoch, SDSS data.
The redshift range of interest  for ground-based Lyman-$\alpha$ BAO studies is restricted to $z>2.15$. In this section, we concentrate on these high-$z$ quasars. 

To extrapolate to fainter targets, the stellar sample is now taken to be a set of random objects in a 7.5~${\rm deg}^2$ region in Stripe 82 around $\alpha_{J2000}=0$. It contains about $1000$ objects per deg$^{2}$ at $g<21$, and $\sim 2500$ at $g<23$. The quasar sample is the one used before  augmented by the new quasars discovered in Stripe 82 using the work presented in this paper (Sec.~\ref{sec:results}).  
We use it to compute the efficiency of quasar recovery in three non-overlapping magnitude bins: $g<21$ (about 11000 quasars), $21<g<22$ (over 5000 quasars) and $22<g<23$  (about 2000 quasars). This sample is highly incomplete for faint objects. Therefore, to compute results integrated up to a given magnitude limit, we weight the efficiencies in each magnitude bin by a theoretical quasar luminosity function (LF) based on~\cite{bib:hopkins} and extrapolated to low luminosities (cf. LSST science book). We also use the quasar LF (corrected by detection efficiencies) to estimate the quasar contamination in the so-called stellar sample. This contamination is negligible in the original sample dominated by stars, but as the threshold on $y_{NN}$ increases, actual quasars contained in the ``stellar" sample begin to dominate the set of selected objects. To compute the rejection levels, their contribution is thus estimated and removed. We estimate the systematic uncertainty on the stellar rejection due to this correction to be of order 1\%.

\begin{figure}[h]
\begin{center}\epsfig{figure=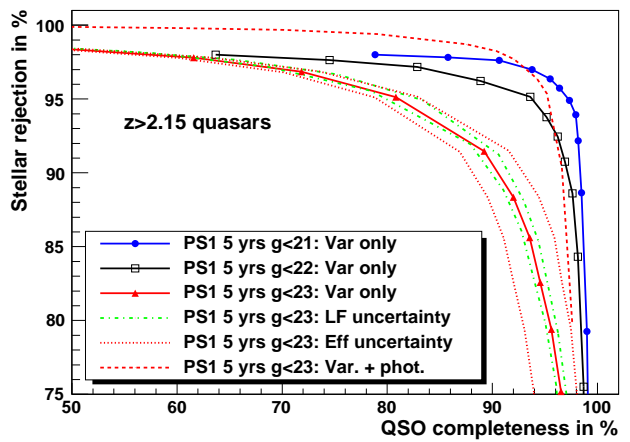,width = \columnwidth} 
\caption[]{Stellar rejection vs. $z>2.15$ quasar completeness for five years of Pan-STARRS simulated data. The depth varies from $g<21$ (blue dots) to $g<23$ (red triangles). The green dot-dashed lines show the effect of a factor of 2 uncertainty on the quasar LF for $22<g<23$. The red dotted lines illustrate the effect of a $\pm 5$\% change on the quasar recovery efficiency. The red dashed line indicates the expected improvement by combining variability and color selections. From left to right, points correspond to $y_{\rm NN}=0.95$, 0.92, 0.87, 0.8, 0.7, 0.6, 0.5, 0.4, 0.3, 0.2, 0.1 and 0.} 
\label{fig:PS1_bigboss}
\end{center}
\end{figure}

Fig.~\ref{fig:PS1_bigboss} shows the stellar rejection $R$ as a function of quasar completeness $C$  for high-$z$ quasars. 
At 80\% quasar completeness (respectively 90\%), the stellar rejection decreases by $\sim 3\%$ (resp. 8\%) when changing the limit from $g<21$ to $g<23$. 

To study the impact of the uncertainty on the quasar LF, we varied the LF in the $22<g<23$ magnitude bin by a factor of 2 either way. As shown in Fig.~\ref{fig:PS1_bigboss} (green dot-dashed lines), even such a large change of the LF has little impact on the results (1\% at most). This is because the quasar recovery efficiency decreases only moderately with increasing magnitude, in particular at large quasar completeness (for $y_{\rm NN}$ near 0.6 or less).  

The uncertainty on the $g<23$ curve is dominated by the uncertainty on the recovery efficiency for quasars in the $22<g<23$ magnitude bin. In this study,  it is estimated at a mean magnitude $g\sim22.3$, lower than what is expected from the quasar LF.  The impact of a 5\% uncertainty on this efficiency is shown in the figure as the red dotted curves: after integration over the full magnitude range, a 5\% change in the recovery efficiency shifts the $g<23$ curve by about 2\%.

The stellar rejection for a given completeness of high-redshift quasars can be improved significantly by combining variability and photometric criteria. With a similar approach as what was done for BOSS on Stripe 82, we define main and extreme variability samples using photometric information from BOSS single-epoch data. The only photometric cut for the main sample is ${\cal P}_{\rm ED}>10^{-3}$, where ${\cal P}_{\rm ED}$ is the probability of extreme deconvolution defined in~\cite{bib:Bovy11}.   This cut rejects $4\%$ of the high-$z$ known quasars. About half of these can be recovered with the extreme variability sample, defined by $y_{NN}>0.95$ and  loose  photometric cuts similar to those applied on Stripe 82 (Sec.~\ref{sec:ancillarysel}). The resulting performance is shown in Fig.~\ref{fig:PS1_bigboss} as the upper red dashed line (for all objects up to $g<23$). 
Considering the $g<23$ curve, relevant to future surveys, we obtain a stellar rejection $R = 99\%$ for a quasar completeness $C=80\%$, and $R=98\%$ for $C=90\%$. Variability alone would have yielded instead $R= 95\%$ and $R= 90\%$ respectively in the same $z>2.15$ redshift range. 
In addition, the photometric selection is optimized for the rejection of low-$z$ quasars, whereas variability is not. 


Although the variability method cannot lead to results as good for the sparser data of Pan-STARRS (13 to 21 measurements in four bands) or PTF (21 measurements in one band) as for the SDSS data on Stripe 82 ($\sim$50 measurements  in five bands), it can still contribute significantly to quasar selection. Used in addition to a color selection, as was done with BOSS for Stripe 82, even with a single epoch in SDSS (for areas other than Stripe 82), it results in much improved selections than what color-selection alone can achieve.

\section{Conclusions}\label{sec:conclusions}
We have designed a method that characterizes light curve variability  in order to discriminate quasars from both non-variable and variable stars. A Neural Network was implemented to yield an estimator of quasar likelihood derived from these variability parameters.

The method has been applied in conjunction with a loose color-based preselection to define a list of 31~$\rm deg^{-2}$ targets in Stripe 82 for which spectra were taken with BOSS. The performance of this selection on quasars at redshift above 2.15 can be quantified by  a purity of 72\% and a completeness of 84\%. This represents a significant improvement over traditional fully color-based selections which seldom obtained a purity in excess of 40\%.  

A second study was dedicated to the objects exhibiting an extreme quasar-like variability. An additional 3~deg$^{-2}$ targets were selected on the following criteria: the objects had to be excluded from the previous sample (i.e. did not have favorable colors according to quasar standards), and had a very high value of the output of the variability NN. Half of the selected objects proved to be high redshift quasars and 40\% low redshift quasars. This program thus increased further the completeness of the quasar selection, reaching the unprecedented value of 90\% total on average over Stripe 82. 

Combining the above two programs allowed BOSS to obtain a density of $z>2.15$ quasars in Stripe 82, all selected through their variability, of 24.0~deg$^{-2}$, with only $\sim$35~deg$^{-2}$ fibers dedicated to their identification.

The method developed here was also applied to ersatz data from Palomar Transient Factory or from Pan-STARRS to determine the performance that can be achieved for future target selections of quasars over about 10,000~deg$^{-2}$ of the sky.

\begin{acknowledgements}
Funding for SDSS-III has been provided by the Alfred P. Sloan Foundation, the Participating Institutions, the National Science Foundation, and the U.S. Department of Energy. The
SDSS-III web site is http://www.sdss3.org/.\\
SDSS-III is managed by the Astrophysical Research Consortium for the Participating Institutions of the SDSS-III Collaboration including the University of Arizona, the Brazilian Participation Group, Brookhaven National Laboratory, University of Cambridge, University of Florida, the French Participation Group, the German Participation Group, the Instituto de Astrofisica de Canarias, the Michigan State/Notre Dame/JINA Participation Group, Johns Hopkins University, Lawrence Berkeley National Laboratory, Max Planck Institute for Astrophysics, New Mexico State University, New York University, the Ohio State University, the Penn State University, University of Portsmouth, Princeton University, University of Tokyo, the University of Utah, Vanderbilt University, University of Virginia, University of Washington, and Yale University.\\
ES is supported by grant DE-AC02-98CH10886. The BOSS French Participation Group is supported by Agence Nationale de la Recherche under grant ANR-08-BLAN-0222.

\end{acknowledgements}

\end{document}